\documentclass[apj,iop]{emulateapj2}
\slugcomment{Accepted to ApJ}

\usepackage{natbib}
\usepackage{epsf}
\usepackage{graphicx}
\usepackage{xspace}    
\usepackage{microtype} 
\usepackage{amsmath}   
\usepackage{amssymb}
\usepackage{float}

\begin{document}

\newcommand{\um}{\ensuremath{\mu\mathrm{m}}\xspace}
\newcommand{\uJy}{\ensuremath{\mu\mathrm{Jy}}\xspace}

\newcommand{\degrees}{$^{\circ}$}
\newcommand{\degsq}{\ensuremath{\mathrm{deg}^2}}
\newcommand{\ssfr}{\ensuremath{\mathrm{sSFR}}\xspace}
\newcommand{\ssfrunit}{\ensuremath{\mathrm{yr}^{-1}}}

\def\mpch {$h^{-1}$ Mpc} 
\def\kpch {$h^{-1}$ kpc} 
\def\kms {km s$^{-1}$} 
\def\lcdm {$\Lambda$CDM } 
\newcommand{\xisp}{\ensuremath{\xi(r_p,\,\pi)}}
\newcommand{\xir}{\ensuremath{\xi(r)}\xspace}
\renewcommand{\wp}{\ensuremath{w_p}\xspace}
\newcommand{\wprp}{\ensuremath{w_p(r_p)}\xspace}
\newcommand{\rp}{\ensuremath{r_p}\xspace}
\newcommand{\pimax}{\ensuremath{\pi_\textrm{max}}}
\newcommand{\hMpc}{\ensuremath{\,h^{-1}\, \textrm{Mpc}}}
\newcommand{\wGG}{\ensuremath{w_\textrm{GG}}}
\newcommand{\wGT}{\ensuremath{w_\textrm{GT}}}
\newcommand{\wTT}{\ensuremath{w_\textrm{TT}}}
\newcommand{\rpvalue}[2]{\ensuremath{\rp\,{#1}\,{#2}\hMpc}\xspace}
\def\xis {$\xi(s)$}
\def\rr {$r_0$}
\def\ggg {$\gamma$}
\def\etal {et al.}
\def\kt{\tilde{k}}
\def\mpc{{\rm Mpc}}
\def\zmax{Z_{\rm max}}

\def\msunyr{\hbox{$M_{\odot}~{\rm yr}^{-1}$}}
\def\deg{\hbox{$^{\circ}$}}

\newcommand{\mass}{\ensuremath{\mathrm{M_*}}}
\newcommand{\mstar}{\ensuremath{\mathcal{M}_*}}
\newcommand{\msun}{\ensuremath{M_{\sun}}}
\newcommand{\logmass}{\ensuremath{\log\,(\mass/\msun)}}

\newcommand{\sfr}{\ensuremath{\psi}}
\newcommand{\sfrm}{{\rm SFR}/\ensuremath{\mass}}
\newcommand{\sfrunit}{\ensuremath{\mathcal{M}_{\sun}~\textrm{yr}^{-1}}}
\newcommand{\iSEDfit}{\texttt{iSEDfit}\xspace}
\newcommand{\arcsecsq}{\ensuremath{\mathrm{arcsec}^2}\xspace}

\title{PRIMUS+DEEP2: The Dependence of Galaxy Clustering on Stellar Mass and Specific Star Formation Rate at $0.2<z<1.2$}

\author{
Alison L. Coil\altaffilmark{1}, 
Alexander J. Mendez\altaffilmark{2}, 
Daniel J. Eisenstein\altaffilmark{3}, 
John Moustakas\altaffilmark{4}
}
\altaffiltext{1}{Center for Astrophysics and Space Sciences,
  Department of Physics, University of California, 9500 Gilman Dr., La
  Jolla, CA 92093}
\altaffiltext{2}{Department of Physics and Astronomy, The Johns
  Hopkins University, 3400 North Charles Street, Baltimore, MD 21218} 
\altaffiltext{3}{Harvard-Smithsonian Center for Astrophysics, 60
  Garden Street, Cambridge, MA 02138}
\altaffiltext{4}{Department of Physics and Astronomy, Siena College,
  515 Loudon Road, Loudonville, NY 12211}

\begin{abstract}
We present results on the clustering properties of galaxies 
as a function of both stellar mass and specific star formation rate (sSFR) using data 
from the PRIMUS and DEEP2 galaxy redshift surveys spanning $0.2 < z < 1.2$.
We use spectroscopic redshifts of over 100,000 galaxies covering an
area of 7.2 deg$^2$ over five separate fields on the sky, from which
we calculate cosmic variance errors.  
We find that the galaxy clustering amplitude is as strong of a function of sSFR 
as of stellar mass, and that at a given sSFR, it does not
significantly depend on
stellar mass within the range probed here.  
We further find that within the star-forming population and
at a given stellar mass, galaxies 
above the main sequence of star formation with higher sSFR are less clustered than galaxies 
below the main sequence with lower sSFR.  We also find that within the quiescent 
population, galaxies with higher sSFR are less clustered than galaxies with 
lower sSFR, at a given stellar mass.  We show that the galaxy clustering amplitude 
smoothly increases with both increasing stellar mass and decreasing sSFR, implying 
that galaxies likely evolve {\it across} the main sequence, not only along it,
before galaxies eventually become quiescent.  These results imply that the stellar 
mass to halo mass relation, which connects galaxies to dark matter halos, 
likely depends on sSFR.  
\end{abstract}


\section{Introduction} \label{sec:intro}
Galaxies are thought to form in the centers of dark matter halos,
regions of the Universe that have collapsed under their own gravity.
The observed clustering of galaxies matches well the predicted
clustering of dark matter halos from $\Lambda$CDM cosmological
numerical simulations, using various prescriptions for assigning
galaxies to halos.  However, it is not yet clear exactly how to map
observed galaxies to dark matter halos, as it is not yet known exactly
how galaxies form and evolve within these halos across cosmic time and
how the dark matter halo influences the galaxy and vice versa.

Earlier galaxy clustering papers often quantified in particular the
luminosity-dependence of clustering, generally finding that the
brightest galaxies are more clustered than fainter
galaxies, with a sharp rise in the clustering amplitude above $L^*$ 
\citep[e.g.,][]{Alimi88, Benoist96, Norberg01}. \nocite{Alimi88}
Similar results were found to hold at higher redshift as well, to $z\sim1$,
when the Universe was less than half its current age 
\citep[e.g.,][]{Coil06, Pollo06, Meneux09}.

 As the
observed bimodality in the optical colors of galaxies became
increrasingly apparent \citep[e.g.,][]{Strateva01,Baldry04}, many
authors turned towards measuring the luminosity-dependence of blue,
star-forming and red, quiescent galaxies
separately \citep[e.g.,][]{Norberg02, Hogg03, Coil04, Zehavi05, Meneux06}.  
These papers showed
that at a given luminosity, red galaxies are more clustered than blue,
and that {\it within} each of these two broad galaxy populations, the
brightest galaxies are typically more clustered than fainter galaxies.
Here again these results were found to hold out to $z\sim1$.  
However, it was also discovered at low redshift 
that within the red, quiescent galaxy
population, low luminosity galaxies are highly clustered, likely
reflecting that they tend to be satellite galaxies in massive dark
matter halos hosting galaxy clusters \citep{Berlind05}.  Some authors
choose to split the galaxy population by morphology or spectral type
instead of color, finding similar results, that galaxies with
early-type, elliptical morphologies or early-type spectra 
are more clustered than late-type,
spiral galaxies \citep[e.g.,][]{Loveday95, Madgwick03, Li06, delaTorre11}.

Moving beyond considering the galaxy population as having only two 
general types, \citet{Coil08} used the DEEP2 galaxy redshift survey 
to split the $z\sim1$ galaxy population 
into finer bins in color, showing that the clustering amplitude 
rises {\it within} the blue, star-forming population alone, as the 
color becomes increasingly red.  They did not find any clustering 
difference within the red, quiescent population when split by optical
color.  \citet{Zehavi11} found using SDSS at $z\sim0$ that 
clustering depends on color both within the blue, star-forming 
population and the red, quiescent population.  
Using the PRIMUS galaxy redshift survey at $z\sim0.7$, \citet{Skibba14}
found again that clustering depends on color within the red, quiescent
population (though not within the blue, star-forming population).
These results began to more fully flesh out how galaxy clustering 
depends on the star formation properties of galaxies, beyond a simple
division into star-forming or quiescent, and pointed to how galaxies
must evolve with time in terms of their color (from very blue to 
very red).

More recently, observers and theorists have moved from mapping the 
galaxy population in color-magnitude space to star formation rate (SFR) or
specific SFR (sSFR, defined as the SFR per unit stellar mass) versus 
stellar mass space \citep[e.g.,][ and references therein]{Noeske07, Speagle14}.  
The latter quantities
are more useful parameters as they are tied to physical processes occuring 
within galaxies (converting gas into stars, the growth of a galaxy) 
and are less impacted by dust obscuration.  They are also easier quantities
for theorists to model in cosmological simulations than color and magnitude.
As a result, more recently there has been a lot of work quantifying the 
stellar mass-dependence of galaxy clustering 
\citep[e.g.,][]{Li06,Meneux08,Wake11,Leauthaud12, Marulli13}.  
These papers typically find that the clustering amplitude is a strong 
positive function of stellar mass above $M^*$ and is less dependent at 
lower stellar masses.  This has led to many papers quantifying the 
stellar mass to halo mass relation and its evolution with cosmic time 
\citep[e.g.,][]{Behroozi10, Moster10, Leauthaud11, Durkalec15, Skibba15}.

While there has been substantial work on the stellar mass-dependence of
galaxy clustering, there have been few papers on the SFR or sSFR-dependence, 
either at low or high redshift. 
In a pair of related papers, \citet{Hearin14} and \citet{Watson15} show that
the clustering properties of SDSS galaxies divided into star-forming or 
quiescent at a given stellar mass are very similar whether the galaxy 
subsamples are defined using either optical colors or sSFR.  Essentially, 
as long as the observed bimodality in the galaxy population is used, whether 
the color or sSFR is used to define the bimodality does not matter in terms of
the relative clustering of blue, star-forming galaxies to red, quiescent 
galaxies, perhaps not surprisingly.  

\citet{Li08} use SDSS to compare low and high sSFR samples within the 
star-forming population, and find that on very small scales (less than 
100 kpc) the clustering amplitude is higher for galaxies with higher sSFR.  
This is likely due to galaxy-galaxy tidal interactions.  
\citet{Heinis09} use GALEX imaging of SDSS to investigate both the $NUV-r$ 
and \ssfr dependence of clustering, finding that the clustering amplitude 
increases with decreasing \ssfr or redder color, where they split the star-forming
population into two bins and compare with the quiescent population. 

Other papers that have divided the fully galaxy population more finely 
into multiple
bins in either SFR or sSFR have typically used only angular clustering 
measurements, where spectroscopic redshifts are lacking for individual 
galaxies \citep{Sobral10, Lin12, Dolley14, Kim15}.  These papers, which 
span $z\sim0.2-2.0$, generally find that galaxy subsamples with higher
SFR or lower sSFR have higher clustering amplitudes. 
\citet{Sobral10} measure the angular clustering of H$\alpha$ emitters 
at $z\sim0.8$ and find that clustering amplitude increases steadily with 
H$\alpha$ luminosity (which is a proxy for SFR), even at a fixed K-band 
luminosity (which is a proxy for stellar mass).  
\citet{Dolley14} measure the angular clustering of star-forming galaxies 
over a wide area of 8 deg$^2$, selecting galaxy subsamples based on 
{\it IRAC/MIPS} 24$\micron$ flux.  They find that galaxies with higher 24$\micron$ flux (which is a proxy for SFR) have higher clustering amplitudes, though they do not investigate whether this difference may be accounted for by differences in the mean stellar mass of the samples.
\citet{Kim15} measure the angular clustering of galaxies at $z\sim1$ in 
the UKIDDS DXS survey as a function of stellar mass and sSFR.  They find a 
steady increase in the clustering amplitude with decreasing sSFR, above a
given stellar mass threshold.

\citet{Mostek13} use the DEEP2 galaxy redshift survey at $z\sim1$ to measure
the stellar mass, SFR, and sSFR dependence of galaxy clustering, using 
multiple bins in each physical parameter.  They find that within the 
star-forming population, clustering amplitude increases with increasing 
SFR and decreasing sSFR, though they find no SFR-depdence for quiescent 
galaxies. They investigate whether the SFR-depdence that is observed could 
be due to stellar mass and conclude that much, though not all, of the trend 
could be due to the known correlation between SFR and stellar mass (the 
star-forming ``main sequence'').  They also investigate small-scale 
clustering properties and find a clustering excess for higher sSFR samples 
both within the star-forming and quiescent populations, which they attribute 
to galaxy-galaxy interactions.

 \citet{Mostek13} also find that star-forming galaxies above the ``main 
sequence'' of star formation are less clustered than those below, within 
a given stellar mass range, which points to the possibility of using clustering 
measurements to track the evolution of galaxies in the SFR-stellar mass plane.
However, the DEEP2 sample is not large enough to further divide the galaxy 
population into multiple bins in SFR and stellar mass.  

Here we use data from the PRIMUS and DEEP2 galaxy redshift surveys to study 
the dependence of galaxy clustering on stellar mass and sSFR using a sample 
of over 100,000 spectroscopic redshifts at $0.2 < z < 1.2$.  Our sample
spans a total of five fields, which we use to quantify errors due to cosmic
variance.  We use deep multi-wavelength imaging in our fields to estimate 
stellar masses and sSFRs, from which we create multiple galaxy subsamples
using cuts in both parameters.  We measure cross-correlation functions 
of these galaxy subsamples with all galaxies in our survey at these redshifts,
to better trace the underlying cosmic web and reduce our uncertainties.

This paper is organized as follows.  In \S\ref{sec:data} we present the 
relevant spectroscopic datasets used here and describe our methodology 
for deriving stellar masses and sSFRs.  In \S\ref{sec:samples} we describe
the various galaxy subsamples used in our clustering analysis.  The methods
used to perform the clustering analysis are presented in \S\ref{sec:methods}, 
and our results are given in \S\ref{sec:results}. 
We discuss our results in \S\ref{sec:discussion} and conclude in \S\ref{sec:conclusions}.
Throughout the paper we assume a standard $\Lambda$CDM model with 
$\Omega_m=0.3$, $\Omega_\Lambda=0.7$, and $H_{0}=72$~km s$^{-1}$~Mpc$^{-1}$.

\section{Data} \label{sec:data}
For this study we use data from the PRIMUS \citep{Coil11, Cool13} and DEEP2 \citep{Newman13} galaxy spectroscopic redshifts surveys.  The data used here are taken from five independent fields on the sky, covering a total of 7.2 deg$^2$.  We use the separate fields to quantify the effects of cosmic variance on the clustering properties of galaxies, as described below.  
Here we use data from the CDFS-SWIRE \citep{Lonsdale03}, COSMOS
\citep{Scoville07}, Elais-S1 \citep[ES1][]{Oliver00} and the XMM-LSS
\citep{Pierre04} fields in the PRIMUS survey, as well as the Extended
Groth Strip (EGS) in the DEEP2 survey. 
We describe the PRIMUS and DEEP2 spectroscopic surveys briefly in Sections~\ref{sec:primusdata} and \ref{sec:deep2data}, and in Section~\ref{sec:massdata} we explain the methods we use to estimate stellar masses and SFRs in these datasets.

\subsection{PRIMUS Redshift Survey}\label{sec:primusdata}

We use spectroscopic redshifts from the PRIMUS redshift survey to perform 
our clustering analysis. PRIMUS is currently the
largest faint galaxy redshift survey completed to date.  The full survey 
covers $\sim9\,\degsq$ in a total of seven well-studied fields on the sky with 
multi-wavelength imaging, including X-ray, infrared (IR) and ultravoilet (UV). 
The survey obtained
low-resolution ($\lambda/\Delta\lambda \sim 40$) spectra with the 
IMACS instrument \citep{Bigelow03} on the Magellan-I Baade 6.5 m
telescope, observing $\sim 2,500$ objects simultaneously over an area of 
0.18 $\degsq$. PRIMUS contains a statistically-complete sample of $\sim120,000$
robust spectroscopic redshifts to $i_{\mathrm{AB}}\sim23.5$. 

Redshifts are derived by
fitting a large suite of galaxy, broad-line AGN, and stellar spectral templates
to the low-resolution spectra and optical photometry \citep[see][for
details]{Cool13}. Objects are classified as galaxies, broad-line AGN or stars
depending on the best $\chi^2$ template fit. The PRIMUS spectroscopic 
redshifts have a 
precision of $\sigma_z/(1+z) \sim 0.5\%$. We use robust
($z_\textrm{quality} \ge 3$, see \citet{Coil11}) PRIMUS redshifts between $0.2
< z < 1.2$ in the CDFS-SWIRE, COSMOS, ES1, and XMM-LSS fields. For further
details of the survey design, targeting, and data see \citet{Coil11}; for
details of the data reduction, redshift confidence, and completeness see
\citet{Cool13}.

The PRIMUS survey generally targeted all sources above $i < 22.5$ and
sparse-sampled $22.5 < i < 23$ sources, so that faint galaxy sources at the
flux limit would not dominate the target selection. The targeting weights were 
defined {\it apriori} such that a statistically complete flux-limited
sample could be recreated, by tracking both the ``sparse sampling'' weight and the
``density dependent'' weight of each object. The sparse 
sampling weight accounts for the fraction of sources selected at random in the
0.5 mag interval above the targeting limit in each field; it is therefore a magnitude-dependent weight. In contrast, the {\it density-dependent} 
weight accounts for sources in high density areas on the plane of the 
sky that can not be targeted due to slit collisions and the number of overlapping masks observed \citep[see ][ for more details]{Coil11, Moustakas13}.
From the full PRIMUS sample, only those targets defined as belonging to the ``primary'' sample have these well-defined targeting weights; hence, for our clustering analysis we use only ``primary'' targets.

For the clustering measurements presented here, 
we also include a 
spatially-varying {\it redshift success} weight
to account for changes in the observed redshift success fraction across a field
(i.e., due to differences in observing conditions for different slitmasks). In
the PRIMUS fields we use the \texttt{pixelize} function in
\texttt{Mangle} to create these weights. We
estimate the redshift success fraction by taking the ratio of robust redshift 
sources with $z_{\textrm{quality}} \ge 3$ to all targeted sources in the field, using pixels of size $\sim 36~\arcsecsq$.

\subsection{DEEP2 Redshift Survey}\label{sec:deep2data}

We also use spectroscopic redshifts from the EGS field of the DEEP2
survey \citep{Newman13}.  The DEEP2 survey was
conducted with the DEIMOS spectrograph \citep{Faber03} on the 10m Keck-II
telescope. In the EGS, the DEEP2 survey has measured $\sim17,000$
high-confidence redshifts ($Q \ge 3$, See \citet{Newman13}) to $R_{AB} = 24.1$.
Unlike the other DEEP2 fields, in the EGS there was no photometric redshift
pre-selection of targets; thus all galaxies that could be observed on 
slitmasks to this photometric depth were targeted.
 We use the Data Release 4 (DR4)
catalog\footnote{http://deep.ps.uci.edu/dr4/home.html} and associated window
function from \citep{Newman13}. We use redshifts between $0.2 < z < 1.2$ 
that have a redshift confidence greater than
95\% ($Q \ge 3$). We use the extended optical photometry from
\citet{Matthews13} which contains additional Canada-France-Hawaii Telescope
Legacy Survey (CFHTLS) $ugriz$ and the Sloan Digital Sky Survey (SDSS) $ugriz$
photometry matched to the redshift catalog. K-corrections, absolute $M_B$
magnitudes, and rest-frame colors are derived from K-correct \citep{Blanton07}
from the optical photometry in these fields. 

As in the PRIMUS fields, in the EGS we also include a 
spatially-varying redshift success weight, which 
reflects the probability that a targeted source has a secure
$z_\mathrm{quality} \ge 3$ redshift. For the EGS we calculate this in
$\sim 6~\arcsecsq$ pixels, as the deeper DEEP2 data allows us to use smaller
pixels than in the PRIMUS fields. However, using the average of six adjacent pixels to match
the $\sim 36~\arcsecsq$ pixels used in PRIMUS does not change the resulting
clustering measurements in this field.

In order to perform accurate clustering measurements, we require that all of
the PRIMUS and DEEP2 sources used here are located within the area of each
survey that has a well-understood spatial selection function. This ensures that
any spatially-dependent density differences in the surveys that are due to
target selection or missing data, such as in CCD chip gaps or around bright
stars, as well accounted for In PRIMUS we require that sources fall within the
observed window function area targeted with at least two slitmasks.
Details of the PRIMUS spatial selection function are given in 
\citet{Coil11}, and \citet{Coil04a} and \citet{Newman13} provide details for 
the DEEP2 survey.

\subsection{Stellar Mass and sSFR Estimates}\label{sec:massdata}

We estimate stellar masses and sSFRs by fitting the
spectral energy distributions (SEDs) of our sources with population synthesis
models using \iSEDfit \citep{Moustakas13}. \iSEDfit is a Bayesian fitting code
that compares the observed photometry for each source to a large Monte Carlo
grid of SED models which span a wide range of stellar population parameters, including age, metallicity, and star formation history, to estimate the stellar
mass and SFR of a galaxy. The sSFR is then simply defined as the SFR divided 
by the stellar mass.
We use \iSEDfit results derived from photometry spanning the UV to the
NIR IRAC bands.
We assume a \citet{Chabrier03} initial mass function
from $0.1$ to $100~\mstar$ and use \citet{Bruzual03} stellar population
synthesis models. We assume the following priors to construct the Monte Carlo
grids: uniform stellar metallicity in the range of $0.004 < Z < 0.04$;
\citet{Charlot00} dust attenuation law, with an exponential distribution of
dust, ($0.25 < \gamma < 2.0$); an exponentially declining-$\tau$ ($\phi_{s}(t)
= (\mathcal{M}/\tau) e^{-t/\tau} $) star-formation history (SFH) with $0.01 <
\tau < 5.0$. Stochastic bursts of star formation of varying amplitude,
duration, and onset time are superimposed, allowing for a wide range of
possible star formation histories.
While a delayed-$\tau$ model encompasses both a linearly rising ($t/\tau \ll
1$) and an exponentially declining ($t/\tau \gg 1$) SFH history, we find no
significant SFR or stellar mass offsets or trends using different SFH models
for our sources at $z<1.2$, and we therefore choose to use a simpler model of
an exponentially declining SFH. \iSEDfit marginalizes the full posterior
probability distribution of stellar masses and SFRs over all other parameters
and thus encapsulates both the uncertainties in the observations and the model
parameter degeneracies. For each source we take the median stellar mass and SFR
from the full probability distribution functions as the best estimate of the
stellar mass and SFR. The median uncertainties on the log stellar mass and 
SFR are 0.08 dex and 0.2 dex, respectively.  While the systematic
errors on the stellar mass and SFR estimates may be larger than the
statistical errors, our concern in this paper is how the relative bias
scales with stellar mass and \ssfr. So long as systematic errors in
determining these parameters do not correlate with large-scale density
on scales $>$1 \mpch, then our conclusions are robust to these systematics.

\section{Galaxy Samples} \label{sec:samples}

\begin{figure*}[th]
\epsscale{1.1}
\plotone{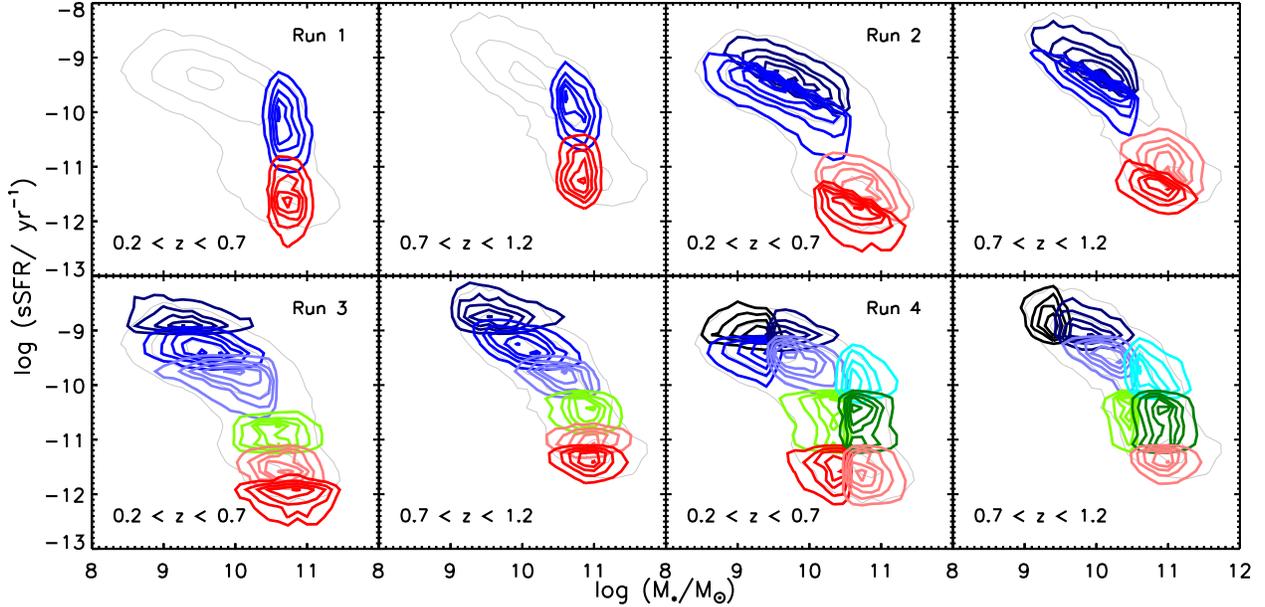}
\caption{\label{fig:samples}
\small Specific star formation rate (\ssfr) versus stellar mass for the various galaxy samples used in this paper.  We divided the full galaxy sample into four ``runs'', each with various galaxy samples defined by cuts in stellar mass and \ssfr, at both lower redshift ($0.2 < z < 0.7$) and higher redshift ($0.7 < z < 1.2$).  Light grey contours show the full galaxy population in the relevant redshift interval, while colored contours show the various galaxy samples used in each of the runs for our clustering analysis.  The justification for the different runs is given in the text. 
}
\end{figure*}

\begin{deluxetable*}{llrrrrrrrr}
\tabletypesize{\footnotesize}
\tablecaption{Galaxy Samples}
\tablehead{
\colhead{Run}&\colhead{Name}&\colhead{N$_{gal}$\tablenotemark{a}}&\colhead{z}&\colhead{}&\colhead{\hspace{-0.3cm}\logmass}\hspace{-1cm}&\colhead{}&\colhead{}&\colhead{\hspace{-0.3cm}log\,(\ssfr/\ssfrunit)}\hspace{-1cm}&\colhead{} \\
\colhead{}&\colhead{}&\colhead{}&\colhead{mean}&\colhead{min}&\colhead{mean}&\colhead{max}&\colhead{min}&\colhead{mean}&\colhead{max} 
}
\startdata
\\
1 & blue-lowz\tablenotemark{b} & 7,418  & 0.51 & 10.50 & 10.71 & 11.00 & $-$11.37 & $-$10.21 & $-$8.25   \\
  & red-lowz     & 6,349 & 0.51  & 10.50 & 10.74 & 11.00 & $-$13.08 & $-$11.61 & $-$10.70  \\
\\
  & blue-highz   & 6,674 & 0.89  & 10.50 & 10.73 & 11.00 & $-$10.77 & $-$9.89 & $-$8.11    \\
  & red-highz    &  5,169 & 0.87  & 10.50 & 10.79 & 11.00 & $-$12.23 &  $-$11.09 & $-$10.16  \\
\\
2 & blue1-lowz   & 21,600 & 0.52  &  8.50 & 9.73  & 10.50 & $-$10.03 & $-$9.26 & $-$7.94    \\
  & blue2-lowz   & 23,795 & 0.41  &  8.50 & 9.59  & 10.50 & $-$11.25 & $-$9.80 & $-$8.75    \\
  & red1-lowz    & 6,797 & 0.56  & 10.10 & 10.76 & 11.60 & $-$12.16 & $-$11.35 & $-$10.59  \\
  & red2-lowz    & 5,641 & 0.42  & 10.10 & 10.64 & 11.60 & $-$13.32 & $-$11.92 & $-$11.26  \\
\\
  & blue1-highz  & 11,087 & 0.89  & 8.70  & 9.91  & 10.50 & $-$9.68  &  $-$9.02 & $-$7.93   \\
  & blue2-highz  & 7,837 & 0.82  & 8.70  & 9.96  & 10.50 &$-$10.62  & $-$9.58  & $-$8.52   \\
  & red1-highz   & 5,372 & 0.92  & 10.10 & 10.97 & 11.60 &$-$11.61  & $-$10.82 & $-$10.05  \\
  & red2-highz   & 4,257 & 0.82  & 10.10 & 10.83 & 11.60 & $-$12.23 & $-$11.41 & $-$10.75  \\
\\
3 & 1-lowz       & 4,934 & 0.53  & 8.50  & 9.26  & 10.50 & $-$9.00  & $-$8.79 & $-$8.00    \\
  & 2-lowz       & 22,744 & 0.47  & 8.50  & 9.53  & 10.50 & $-$9.60  & $-$9.33 & $-$9.00    \\
  & 3-lowz       & 16,271 & 0.44  & 8.50  & 9.91  & 10.50 & $-$10.60 & $-$9.93 & $-$9.60    \\
  & 4-lowz       & 5,437 & 0.51  & 10.00 & 10.61 & 11.50 & $-$11.20 & $-$10.90 & $-$10.60  \\
  & 5-lowz       & 6,817 & 0.52  & 10.00 & 10.67 & 11.50 & $-$11.80 & $-$11.51 & $-$11.20  \\
  & 6-lowz       & 3,824 & 0.39  & 10.00 & 10.78 & 11.50 & $-$12.60 & $-$12.06 & $-$11.80  \\
\\
  & 1-highz      & 3,861 & 0.90  & 9.00  &  9.66 & 11.00 & $-$8.90 & $-$8.66 & $-$8.00     \\
  & 2-highz      & 12,770 & 0.87  & 9.00  & 10.04 & 11.00 & $-$9.60 & $-$9.27 & $-$8.90     \\
  & 3-highz      & 6,914 & 0.87  & 9.50  & 10.51 & 11.00 & $-$10.20 & $-$9.85 & $-$9.60    \\
  & 4-highz      & 4,888 & 0.88  & 10.20 & 10.88 & 11.70 & $-$10.80 & $-$10.49 &$-$10.20   \\
  & 5-highz      & 3,337 & 0.89  & 10.20 & 10.93 & 11.70 & $-$11.20 & $-$11.00 & $-$10.80  \\
  & 6-highz      & 4,109 & 0.84  & 10.20 & 10.93 & 11.70 & $-$11.80 & $-$11.42 & $-$11.20  \\
\\
4 & 1-lowz       &  7,067 & 0.49  & 8.50 & 9.12   & 9.50  & $-$9.20  & $-$8.95  & $-$8.20   \\
  & 2-lowz       & 10,577 & 0.38  & 8.50 & 9.18   & 9.50  & $-$10.20 & $-$9.48  & $-$9.20   \\
  & 3-lowz       &  3,494 & 0.56  & 9.50 & 9.78   & 10.50 & $-$9.20  & $-$9.02  & $-$8.20   \\
  & 4-lowz       & 19,817 & 0.49  & 9.50 & 9.96   & 10.50 & $-$10.20 & $-$9.65  & $-$9.20   \\
  & 5-lowz       &  5,698 & 0.45  & 9.50 & 10.15  & 10.50 & $-$11.20 & $-$10.65 & $-$10.20  \\
  & 6-lowz       &  3,618 & 0.42  & 9.50 & 10.20  & 10.50 & $-$12.20 & $-$11.59 & $-$11.20  \\
  & 7-lowz       &  3,870 & 0.53  & 10.50 & 10.74 & 11.50 & $-$10.20 & $-$9.85  & $-$9.20   \\
  & 8-lowz       &  5,875 & 0.53  & 10.50 & 10.80 & 11.50 & $-$11.20 & $-$10.68 & $-$10.20  \\
  & 9-lowz       &  6,913 & 0.51  & 10.50 & 10.86 & 11.50 & $-$12.20 & $-$11.67 & $-$11.20  \\
\\                
  & 1-highz      &  2,291 & 0.82  & 8.50  & 9.31  & 9.50  & $-$9.20  & $-$8.76  & $-$8.20   \\
  & 2-highz      &  6,232 & 0.90  & 9.50  & 9.89  & 10.50 & $-$9.20  & $-$8.94  & $-$8.20   \\
  & 3-highz      &  9,674 & 0.85  & 9.50  & 10.13 & 10.50 & $-$10.20 & $-$9.53  & $-$9.20   \\
  & 4-highz      &   944 & 0.79  & 9.50  & 10.36 & 10.50 & $-$11.20 & $-$10.61 & $-$10.20  \\
  & 5-highz      &  5,964 & 0.91  & 10.50 & 10.80 & 11.50 & $-$10.20 & $-$9.80  & $-$9.20   \\
  & 6-highz      &  7,295 & 0.89  & 10.50 & 10.94 & 11.50 & $-$11.20 & $-$10.70 & $-$10.20  \\
  & 7-highz      &  3,949 & 0.84  & 10.50 & 10.95 & 11.50 & $-$12.10 & $-$11.44 & $-$11.20  \\
\enddata
\tablenotetext{a}{This is the weighted number of galaxies in each sample; weights are discussed in Section 2.}
\tablenotetext{b}{The ``low-z'' samples span $0.2 < z < 0.7$, while the ``high-z'' samples span $0.7 < z < 1.2$.}
\label{tab:samples}
\end{deluxetable*}

The goal of this paper is to quantify the dependence of galaxy clustering at intermediate redshift on stellar mass and \ssfr.  To facilitate this, we created various galaxy samples from the full galaxy population, which is defined as all galaxies with robust redshifts (as described above) at $0.2 < z < 1.2$.  
We create galaxy samples four times, which we call four ``runs'', 
using different cuts in stellar mass and \ssfr for each run.  
We always create galaxy samples in two redshift intervals for each run: $0.2 < z < 0.7$ and $0.7 < z < 1.2$. 

We identify star-forming and quiescent galaxies based on their location in the SFR versus stellar mass plane, using an evolving linear relation that traces the minimum of the bimodal galaxy distribution in PRIMUS:

\begin{equation}
\log\,({\rm SFR}) = -1.29 + 0.65\,(\log\,\mass - 10) + 1.33\,(z - 0.1)
\label{eq:SFR}
\end{equation}

\noindent where SFR has units of $\sfrunit$ and $\mass$ has units of $\msun$.
The slope of this line is defined by the slope of the star forming main sequence \citep[e.g.,][]{Noeske07} as measured in the PRIMUS dataset using \iSEDfit SFR and stellar mass estimates. 
Each galaxy is classified as star-forming or quiescent based on whether it lies above or below the cut defined by Equation~\ref{eq:SFR}, evaluated at the redshift of the galaxy.  Fig.~2 of \citet{Berti16} shows the distribution of PRIMUS galaxies in the SFR--stellar mass plane in bins of redshift, along with the location of this cut. 

The details of the various samples created here 
for each run are given in Table~\ref{tab:samples}, and the location of the galaxy samples in the \ssfr versus stellar mass plane are shown in Figure~\ref{fig:samples}.
The $i$ and $R$-band selection limits used in the PRIMUS and DEEP2 redshift surveys correspond to $\sim4000$\AA \ restframe selection at $z\sim0.7$, such that galaxies with higher \ssfr (i.e., star-forming galaxies) are included in our sample at lower stellar masses than quiescent galaxies with lower \ssfr.  This can clearly be seen by the lack of galaxies in the lower left regions in Fig~\ref{fig:samples}.  

For the samples described below, the lower redshift samples in the first run are stellar mass-limited, in that all galaxies in these samples are above the stellar mass completeness limits of the PRIMUS survey \citep{Moustakas13}.  This facilitates comparisons with theoretical models that required stellar mass-complete samples.  The rest of the samples are flux-limited.  This is required in order to probe a wide enough range in both stellar mass and \ssfr for the purposes of investigating the joint clustering dependence on these properties.  These samples are useful for quantifying the relative clustering dependence but should not be interpreted as being complete to all stellar masses and \ssfr values at the mean values of each sample.

For the first run, we are interested in comparing star-forming and quiescent galaxies at the same stellar mass.  Therefore we 
restrict the stellar mass range to $10.5 < \logmass < 11.0$ and create two galaxy samples in each redshift range using Equation~\ref{eq:SFR}.  These samples are shown in the upper left panels of Fig.~\ref{fig:samples} labelled ``Run 1'', and we will additionally refer to this run below in the text as the ``star-forming/quiescent split'' run.  This run allows us to compare star-forming and quiesent galaxies at similar stellar masses (the mean stellar masses of these samples differ by only $\sim$0.03--0.06 dex, as seen in Table~\ref{tab:samples}).  As stated
above, the lower redshift samples in this run are stellar mass-complete.

For the second run, we are interested in comparing galaxies of higher or lower sSFR {\it within} the star-forming and quiescent populations separately.  In this run we are not concerned with comparing the clustering properties of star-forming with those of quiescent galaxies, therefore we use wider stellar mass ranges than in run 1, and we did not require the same stellar mass ranges for the star-forming and quiescent populations.  Here we choose samples that effectively split the star-forming population into those galaxies above and below the main sequence of star formation, and within the quiescent population into those galaxies that are higher \ssfr than those that are more quiescent.  We refer to this run below in the text as the ``main sequence split'' run. Within the star-forming population we require the stellar mass to be within $8.5 < \logmass < 10.5$ and use the following cuts:
\begin{eqnarray}
\log\,(\ssfr) + 0.65\,(\log\,\mass-10) > -9.715 \\
\log\,(\ssfr) + 0.65\,(\log\,\mass-10) > -9.365
\end{eqnarray}
at $0.2 < z < 0.7$ and $0.7 < z < 1.2$, respectively.
Within the quiescent population we restrict the mass range to $10.1 < \logmass < 11.6$ and use the following cuts:
\begin{eqnarray}
\log\,(\ssfr) + 0.65\,(\log\,\mass-10.5) > -11.512 \\
\log\,(\ssfr) + 0.65\,(\log\,\mass-10.5) > -10.937
\end{eqnarray}
at $0.2 < z < 0.7$ and $0.7 < z < 1.2$, respectively.

We also ran but do not show here more simple divisions of the star-forming and quiescent populations using a strict cut in \ssfr.  We find very similar results to using the cuts above that include the stellar mass-dependent tilt in the star-forming main sequence (which is derived here for our PRIMUS and DEEP2 sample, using our stellar mass and \ssfr estimates). 

For the third run, we use strict cuts in \ssfr \ to split the galaxy sample in bins of \ssfr, allowing for different stellar mass ranges in the star-forming and quiescent populations. We call this run the ``sSFR cuts'' run.  For this run at $0.2 < z < 0.7$ within the star-forming population we restrict the stellar mass range to be within $8.5 < \logmass < 10.5$ and use cuts in log\,(\ssfr/\ssfrunit) $= -$9.0 and $-$9.6.  
Within the quiescent population we restrict the stellar mass range to be within $10.0 < \logmass < 11.5$ and use cuts in log\,(\ssfr/\ssfrunit) $= -$11.2 and $-$11.8.  
For the higher redshift range, $0.7 < z < 1.2$, within the star-forming population we restrict the stellar mass range to be within $9.0 < \logmass < 11.0$ and use cuts in log\,(\ssfr/\ssfrunit) $= -$9.0 and $-$9.6.  
Within the quiescent population we restrict the stellar mass range to be within $10.2 < \logmass < 11.7$ and use cuts in log\,(\ssfr/\ssfrunit) $= -$10.8 and $-$11.2.  
This run effectively allows us to divide both the star-forming and quiescent populations in three samples each, based on \ssfr.

For the fourth and last run, we are interested in creating samples with either the same \ssfr and different stellar mass or the same stellar mass and different \ssfr, to investigate the dependence of galaxy clustering on one parameter while holding the other parameter fixed.  This run is used solely when
measuring the relative bias between galaxy samples in section 5.2 below. 
At $0.2 < z < 0.7$ we define a total of nine samples and at $0.7 < z < 1.2$ we define a total of seven samples based on stellar mass cuts at \logmass = 9.5, 10.5, and 11.5 and log\,(\ssfr/\ssfrunit) = 9.2, 10.2, and 11.2.  The various samples are shown in the lower right panels of Fig.~\ref{fig:samples} and the parameters of each sample are listed in Table~\ref{tab:samples}.

Additionally, as dicussed below we employ both auto- and cross-correlation function measurements in our analysis.  The advantange of cross-correlation measurements is that it allows us to use the full galaxy population, without making cuts on stellar mass or \ssfr, to trace the cosmic web of large-scale structure with more precision than is possible using smaller galaxy samples.  For these cross-correlation function measurements, we create a ``tracer'' galaxy sample which is simply defined as all galaxies in the full sample in the relevant redshift range. 
The ``tracer'' sample contains 69,720 galaxies at $0.2 < z < 0.7$ and 
37,721 galaxies at $0.7 < z < 1.2$.
We then cross-correlate this ``tracer'' galaxy sample with the various samples defined above.

\section{Methods} \label{sec:methods}
We measure the spatial distribution of galaxies using the two-point correlation
function, which quantifies the excess probability above Poisson of finding two
sources with a given physical separation. While most galaxy clustering 
studies measure the auto-correlation function (ACF) of the galaxy subsample 
of interest, here we measure both the ACF directly and 
also measure the cross-correlation function (CCF) of the galaxy 
subsample of interest with a tracer galaxy sample, from which we then
infer the ACF of the subsample of interest alone. 
The main advantage of this method is that it reduces
the error bars on the ACF for small galaxy subsamples, as the tracer sample
has a much higher space density and is used to more fully trace the 
underlying large-scale structure.
Details of how we perform these measurements and measure both absolute and 
relative biases are given below.

\subsection{Measuring the Two-Point Correlation Function}

The two-point correlation function \xir \ is defined as a measure of
the excess probability $dP$ (above that for an unclustered distribution) 
of finding a galaxy in a volume
element $dV$ at a separation $r$ from another randomly-chosen galaxy,
\begin{equation}
dP = n [1+\xi(r)] dV,
\end{equation}
where $n$ is the mean number density of the galaxy sample in question
\citep{Peebles80}.

For each galaxy subsample we construct a randomly-distributed catalog 
with the same overall sky coverage and redshift distribution as the 
data. The random catalog includes information on the redshift success fraction,
as discussed above.
We then measure the two-point correlation function using the
\citet{Landy93} estimator,
\begin{equation}
\xi=\frac{1}{RR}\left[DD \left(\frac{n_R}{n_D}
\right)^2-2DR\left(\frac{n_R}{n_D} \right)+RR\right],
\end{equation}
where $DD, DR$, and $RR$ are weighted counts of pairs of galaxies 
(as a function of separation) in the data--data,
data--random, and random--random catalogs, and $n_D$ and $n_R$ are the
mean weighted number densities of galaxies in the data and random catalogs.
Weights are used to account for target selection in the PRIMUS sample (see
Section~\ref{sec:data}); by applying these weights we are able to create
a statistically-complete sample that is not subject to spatial biases. In the
DEEP2 fields the weights are included in the spatial selection function which
we use to generate the random catalogs, such that galaxies have unity weight.
In order to determine the radial function of the random catalogs, we
used a high-pass filter in combination with boxcar smoothing 
of the redshift distribution of the galaxies in each field.  This
preserves the shape due to the selection function of the survey
while removing deviations due to large-scale structure.

The ACF measures the clustering of a single sample, where
the two sources are from the same sample, while the CCF measures the clustering
of one type of source, taken from one sample, around that of another type of
source, taken from a second sample. Here we measure the CCF of the galaxy
subsample of interest with the ``tracer'' galaxy sample, which is all 
galaxies with robust redshifts in the redshift range of interest.  
To measure the CCF between two galaxy samples,
we measure the observed number of galaxies from a given sample around
each galaxy in the other sample as a function of distance, divided by
the expected number of galaxies for a random distribution.  
 We use the \citet{Davis83} estimator:
\begin{equation} \label{eqn:davis}
  \xi(r) = \frac{D_1 D_2(r)}{D_1 R(r)} - 1 
\end{equation}
\noindent where $D_1 D_2(r)$ is the sum of the weighted pairs of galaxies between the two samples and $D_1 R(r)$
is the sum of the weighted galaxy-random pairs, both as a function of separation.
Here again weights are used to account for target selection in the
PRIMUS survey and the spatial selection function in the DEEP2 survey.

Peculiar velocities distort \xir\ measurements along the line of sight. 
We therefore measure \xir\ in two dimensions,
\xisp, where \rp\ is the separation perpendicular to the line of sight, which
is unaffected by peculiar velocities, and $\pi$ is the separation along the
line of sight. Integrating \xisp\ along the $\pi$ dimension leads to a statistic
that is independent of redshift space distortions, the projected correlation
function:
\begin{eqnarray}
  \wprp &=& \; 2 \int_0^\infty d\pi\; \xisp \\
          &\approx& \; 2 \int_0^{\pimax} d\pi\; \xisp
\end{eqnarray}
\noindent where \pimax\ is the maximum $\pi$ separation to which we integrate.
As the signal to noise of \xisp\ declines quickly for large values of $\pi$, we
measure the projected correlation function by integrating to a given \pimax\ to
limit shot noise. We use a limit of $\pimax\,=\,40\hMpc$ in both the PRIMUS 
and DEEP2 surveys.

\subsection{Jackknife Error Estimation}\label{sec:jackknife}

We estimate the uncertainty in our measurements using jackknife resampling of
the data. For reasonably large surveys like PRIMUS and DEEP2 
jackknife errors are generally similar to the
cosmic variance errors in \wp derived from simulated mock catalogs \citep[e.g.,][]{Coil08, Skibba14}. We use 11 jackknife samples across our five fields, where we have
spatially subdivided the larger fields (CDFS-SWIRE and XMM-LSS) into two or 
more subfields along lines of constant RA and declination such that the
resulting subsamples probe roughly similar volumes and cover an area on the sky
approximately equal to $\sim$1 \degsq.

The uncertainty in \wp is estimated by calculating the projected correlation
function using each jackknife sample. From this collection of \wp estimates we
calculate the variance in the projected correlation function,
\begin{eqnarray}
    \sigma_{\wp}^{2}(\rp) = \frac{N-1}{N}\sum_{j}^{N}(\wprp - 
                                                     \hat{w}_j(\rp))^{2},
\end{eqnarray}
\noindent where the $N$ is the number of jackknife samples, $j$ indexes each
jackknife sample, and $\hat{w}_j(\rp)$ is the projected correlation function
computed for a given jackknife sample. By measuring the projected correlation
function using multiple fields across the sky, the jackknife resampling 
estimates the uncertainty on our measurements due to cosmic variance.

\subsection{Inferring the Auto-Correlation Function}\label{sec:auto}

In addition to directly measuring the ACF of the various galaxy subsamples, 
we also infer the ACF of these subsamples using the measured CCF with the 
tracer sample. To do this, we also measure the ACF of the tracer sample 
in the same volume as the galaxy subsample of interest. 
We integrate all ACFs and CCFs to the same \pimax \ limit. 
We then infer the ACF of the galaxy sumbsample of interest using 
\begin{equation}
  \wGG(\rp) = \frac{\wGT^2(\rp)}{\wTT(\rp)}
\end{equation}
\noindent where \wGG\ is the projected ACF of the galaxy subsample of interest, \wGT\ is the projected galaxy-tracer CCF,
and \wTT\ is the projected tracer ACF. Implicit is the assumption that the
spatial distributions of the galaxies of interest and the tracer 
galaxies are linearly related to the
underlying dark matter spatial distribution (i.e., that the bias is linear, see
Section~\ref{sec:bias} below) and that galaxies of interest and the tracer galaxies are well mixed within dark matter halos.  To validate this assumption, below we compare the directly-measured ACF of both star-forming and quiescent galaxies with the ACF inferred from the CCF and find excellent agreement on both small and large projected scales, well within the errors.

\subsection{Power Law Fits}\label{sec:powerlaw}
The two-point correlation function can roughly be fit by a power law, with 
$\xir=(r/r_0)^\gamma$, where the scale factor $r_0$ is the scale at which 
there is unity excess probability and $\xi=1$. 
An analytic form can then be fit to \wprp:
\begin{eqnarray}
  \wprp = \, \rp \left(\frac{r_0}{\rp}\right)^\gamma 
                 \frac{\Gamma(\tfrac{1}{2})\Gamma(\tfrac{\gamma-1}{2})}
                                          {\Gamma(\tfrac{\gamma}{2})}
\end{eqnarray}
\noindent where $\Gamma$ is the Gamma function. We fit this analytic function
to our clustering measurements in the approximately linear regime of
$1 < $ \rp$ < 10$ \mpch.
On larger scales the size of our fields limits 
the number of pair counts, which artificially lowers the measured 
correlation function and leads to large statistical fluctuations. 
While power law fits can also be performed on smaller scales, here we present
power law fits only on scales of $1 < $ \rp$ < 10$ \mpch \ and present bias
analyses on both small and large scales.

\subsection{Absolute and Relative Bias Measurements}\label{sec:bias}

We use the measured projected correlation function to estimate
the absolute bias, or dark matter bias, of the various galaxy subsamples. 
The absolute bias $b$ measures the relative
clustering strength of the galaxy subsample to that of dark matter particles. We
estimate this bias at the median redshift of each galaxy subsample using the publicly
available code of \citet{Smith03}. We integrate the dark matter correlation
function to a $\pimax\,=\,40\,\hMpc$ and then calculate the bias as 
\begin{equation}
  b = \sqrt{\frac{w_\textrm{G}}{w_\textrm{DM}}}
\end{equation}
\noindent where $w_\textrm{G}$ is the galaxy ACF and $w_\textrm{DM}$ is the dark
matter ACF on scales of $1 < \rp < 10$ \mpch.  We determine
$w_\textrm{DM}$  at the mean redshift of the relevant galaxy sample.
When comparing the clustering of
different samples --- particularly with other published papers --- 
it is useful to compare the bias values instead of the 
clustering scale lengths, as the bias accounts
for differences in the median redshift of each sample and further 
does not assume that $\xi$ is a power law.  The clustering scale length is also
covariant with the slope of the power law, such that ideally the slopes should be 
fixed when comparing results for different samples.  It is therefore preferred 
to compare the bias values.

Additionally, the {\it relative} bias between two galaxy subsamples is defined as
the square root of the ratio of their respective projected correlation
functions. This allows for a simple comparison of the clustering strength of
two samples and is akin to comparing their absolute bias (relative to dark
matter) values. We estimate the relative bias on two scales: $0.1 < $\rp$ < 1$ \mpch
(which we refer to as the ``one-halo'' or ``small-scale'' relative bias) 
and $1 < $\rp$ < 10$ \mpch
(which we refer to as the ``two-halo'' or ``large-scale'' 
relative bias).
We use the ratio of the CCFs to measure the relative bias between two galaxy subsamples.
Below we present the mean and $1\sigma$ uncertainty of the relative bias across
the jackknife samples when comparing two samples.

\section{Results} \label{sec:results}

\begin{figure*}[ht!]
\plotone{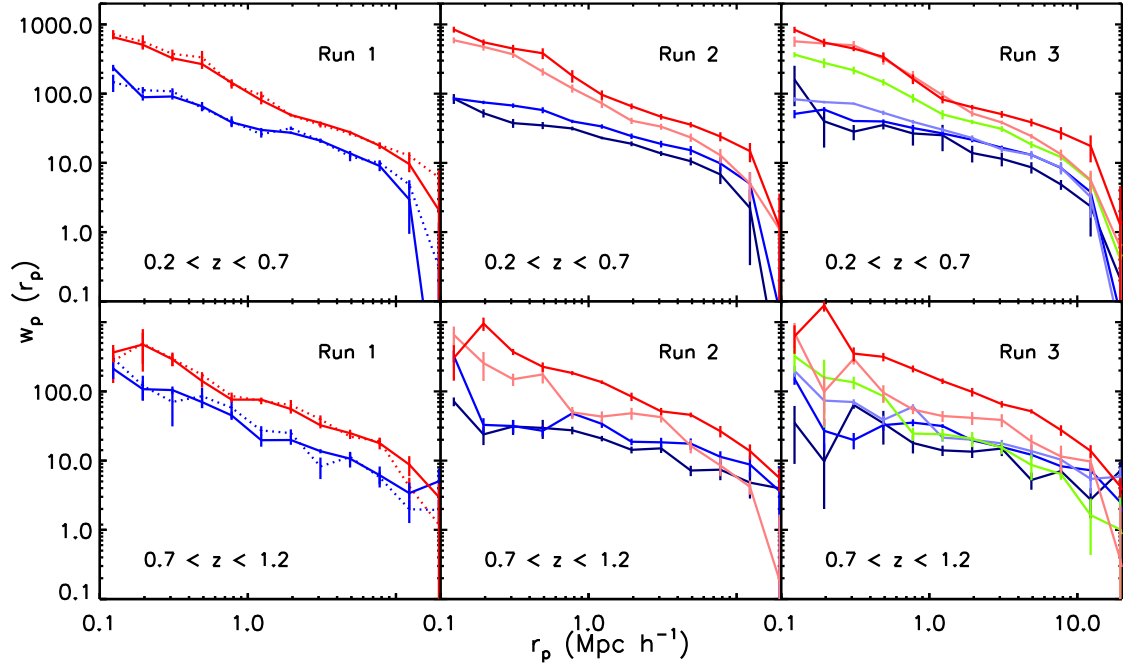}
\caption{\label{fig:wprp}
\small \wprp \ for the galaxy samples in runs 1 
(star-forming/quiescent split), 2 (main sequence split), and 3 (sSFR cuts).  The colors of each sample correspond to the colors shown in Fig.~\ref{fig:samples}. 
The upper row shows \wprp \ for the redshift range $0.2 < z < 0.7$ and the lower row shows \wprp \ for $0.7 < z < 1.2$. In the left column, for run 1, the dotted lines show \wprp \ derived from auto-correlation function measurements, while the solid lines show \wprp \ derived from cross-correlation function measurements.  Given the excellent agreement between them, we utilize \wprp \ derived from cross-correlation function measurements only for all results in this paper. 
}
\end{figure*}

\begin{deluxetable*}{rrrrr}
\tablecaption{\wp Measurements for Run 1 (``Star-forming/Quiescent Split'')}
\tablehead{
\colhead{\rp}&\colhead{blue-lowz}&\colhead{red-lowz}&\colhead{blue-highz}&\colhead{red-highz}
}
\startdata
0.12  & 235.63 (24.73) &  655.07  (46.46)  &   212.64 (62.31) & 364.58  (103.23) \\
0.20  &   88.64  (9.29)  &  508.59  (62.78)  &   107.62 (33.29) & 475.78  (226.77) \\
0.31  &   91.07  (8.94)  &  324.85  (32.43)  &   104.45 (13.16) & 290.76    (58.29) \\
0.49  &   65.70 (10.07) &  266.88  (39.35)  &     70.03 (12.42) & 140.93    (26.38) \\
0.78  &   38.60  (4.47)  &  144.27  (17.54)  &     44.97   (6.29) &   76.02      (6.84) \\
1.23  &   29.93  (2.00)  &    80.25  (10.30)  &     19.72   (3.85) &   75.96      (4.54) \\
1.95  &   27.27  (1.19)  &    49.16    (2.28)  &     19.84   (2.90) &   55.59      (7.84) \\
3.09  &   20.95  (1.49)  &    37.69    (1.94)  &     13.68   (0.94) &   32.37      (3.21) \\
4.90  &   13.71  (1.17)  &    27.70    (1.39)  &     10.67   (2.72) &   24.94      (2.97) \\
7.76  &    9.03  (1.45)   &    17.69    (1.89)  &       6.15   (2.07) &   17.82      (3.29) \\
12.30 &   2.96  (2.02)   &      9.61    (2.28)  &       3.39   (1.41) &     8.77      (2.82) \\
19.50 &   0.00  (0.46)   &      2.08    (3.21)  &       5.02   (2.69) &     2.98      (1.70) \\
\enddata        
\label{tab:wp1}
\end{deluxetable*}

\begin{deluxetable*}{rrrrrrrrr}
\tablecaption{\wp Measurements for Run 2 (``Main Sequence Split'')}
\tablehead{
\colhead{\rp}&\colhead{blue1-lowz}&\colhead{blue2-lowz}&\colhead{red1-lowz}&\colhead{red2-lowz}&\colhead{blue1-highz}&\colhead{blue2-highz}&\colhead{red1-highz}&\colhead{red2-highz}
}
\startdata
0.12  &  84.81 (13.83) &  85.36 (4.98) &  591.45 (57.97) &  843.65 (82.71) &  71.36 (10.34) &  325.09 (65.39) &  655.14 (209.32) &   303.74 (160.99) \\
0.20  &  52.27  (7.09) &  74.84 (4.15) &  474.33 (56.61) &  552.70 (53.36) &  23.89  (7.14) &   32.84  (8.22) &  256.24 (114.53) &   955.32 (207.93) \\
0.31  &  37.56  (5.52) &  67.27 (4.98) &  370.36 (42.74) &  447.45 (58.80) &  31.10  (7.61) &   31.56  (6.72) &  149.96  (30.34) &   372.66  (37.84) \\
0.49  &  34.80  (4.10) &  57.87 (5.78) &  207.14 (25.54) &  383.22 (68.26) &  29.64  (9.09) &   26.93  (4.68) &  175.59  (45.73) &   226.58  (35.44) \\
0.78  &  31.39  (1.96) &  39.77 (2.21) &  119.25 (15.88) &  181.31 (37.13) &  27.40  (3.45) &   48.13  (5.96) &   49.80   (9.84) &   184.03  (12.55) \\
1.23  &  22.92  (0.98) &  33.59 (2.45) &   72.25 (10.32) &   96.36 (15.08) &  20.85  (1.88) &   33.92  (4.23) &   43.28   (8.03) &   135.78   (7.98) \\
1.95  &  18.91  (1.52) &  24.34 (2.24) &   40.65  (4.26) &   65.76  (6.37) &  14.31  (1.62) &   18.73  (2.05) &   48.32   (9.46) &    84.11  (11.65) \\
3.09  &  13.69  (1.10) &  18.77 (2.07) &   33.37  (3.31) &   46.40  (4.38) &  15.01  (2.13) &   18.36  (2.99) &   42.43   (6.92) &    51.47   (6.11) \\
4.90  &  10.57  (1.38) &  15.23 (2.28) &   23.27  (3.16) &   35.62  (3.19) &   7.19  (1.22) &   17.58  (3.31) &   16.32   (3.62) &    45.78   (3.58) \\
7.76  &   6.80  (1.86) &   9.82 (2.22) &   12.88  (3.02) &   24.07  (3.62) &   7.39  (2.18) &   11.34  (2.30) &    8.52   (2.55) &    26.61   (4.98) \\
12.30 &   2.23  (1.90) &   5.02 (2.26) &    5.04  (2.28) &   14.79  (4.58) &   4.81  (1.98) &    8.77  (3.05) &    4.19   (0.97) &    13.74   (3.40) \\
19.50 &   0.01  (0.81) &   0.07 (0.52) &    1.10  (2.49) &    1.19  (2.41) &   3.98  (2.46) &    3.64  (4.67) &    0.19   (1.48) &     5.60   (2.05) \\
\enddata                                      
\label{tab:wp2}
\end{deluxetable*}

In this section we present the two-point correlation functions of the various galaxy samples defined by cuts in stellar mass and \ssfr, along with the dependence of the absolute bias on these parameters.  We also investigate how the relative bias between galaxy samples depends on stellar mass and \ssfr, and show that the dependence on \ssfr is stronger than the dependence on stellar mass.

\subsection{\wprp and Absolute Bias of Galaxy Samples}

Figure~\ref{fig:wprp} shows the two-point correlation function of the galaxy samples for runs 1, 2, and 3.  
For run 1 (``star-forming/quiescent split''), where we divide the galaxy population into star-forming versus quiescent for a limited stellar mass range, we show both the directly-measured ACF (dotted lines) and the inferred ACF derived using the CCF with the tracer galaxy sample (solid lines).  The excellent agreement between these demonstrates that the CCF can be used to robustly recover the ACF.  While the CCF can result in an artificially low \wp measurement on large scales (as seen in the upper left panel of Figure~\ref{fig:wprp}), the resulting bias measurement decreases by only 1\%; this 1\% systematic is well worth the substantially reduced cosmic variance jackknife errors that are derived using the cross-correlation function.
Tables~\ref{tab:wp1} and \ref{tab:wp2} list the \wp measurements for the samples in runs 1 and 2, with the jackknife errors given in parentheses. 
Table~\ref{tab:wprp} lists the power law fits to the \wprp results shown in Fig.~\ref{fig:wprp}, along with the absolute bias of each sample.  We focus in this section on results from runs 1, 2, and 3 for clarity; run 4 is used below in Section~\ref{sec:relbias} where we present relative bias results.

The clustering results from run 1 (``star-forming/quiescent split'') clearly show that at a similar stellar mass, quiescent galaxies are substantially more clustered than star-forming galaxies.  Both the slope and correlation scale length are higher for the quiescent galaxy sample, in both redshift ranges.  
In run 2 (``main sequence split'') we further divide both the star-forming and quiescent populations by \ssfr, or more precisely, whether they are above or below the ``main sequence'' of star formation or a similarly-sloped ridge in the quiescent population, and we find here that, again at a given stellar mass, galaxies above the main sequence are less clustered than galaxies below the main sequence.  Within the quiescent population, galaxies with higher \ssfr are also less clustered than galaxies with lower \ssfr.  The slope of \wprp does not vary substantially within either the star-forming or quiescent populations, but at least at $0.2 < z < 0.7$ where we have smaller error bars, there is a clear change in the slope {\it between} the star-forming and quiescent populations, as seen in run 1.  We do note, however, that the mean stellar mass varies by $\sim1$ dex in run 2 between the star-forming and quiescent populations; the difference is much smaller in run 1.

In run 3 (``sSFR cuts'') we split the full galaxy population into six bins in \ssfr, allowing the mean stellar mass to change as needed to create large galaxy samples (here again the difference is $\sim$1 dex between the star-forming and quiescent populations).  Here we find that both the clustering scale length and slope generally increase with decreasing \ssfr.  As seen in Fig.~\ref{fig:wprp}, within the 
quiescent population in run 3 at $0.2 < z < 0.7$ (upper right panel) there is not a difference in the clustering properties on scales $0.1 < $ \rp $< 1$ \mpch, but there is a difference on scales $1 < $ \rp $ < 10$ \mpch.  
At $0.7 < z < 1.2$ there is a difference within the quiescent population on all scales, though the error bars are larger in our higher redshift bin.

\begin{deluxetable}{llccc}[htb]
\tablecaption{Power-law and Bias Measurements}\tablenotemark{*}
\tablehead{
\colhead{Run}&\colhead{Name}&\colhead{r$_0$}&\colhead{$\gamma$}&\colhead{Bias}
}
\startdata
\\
1 & blue-lowz   & 3.63  $\pm 0.14$  &  1.57  $\pm 0.05$  &  1.23 $\pm 0.08$ \\
  & red-lowz    & 5.96  $\pm 0.20$  &  1.82  $\pm 0.11$  &  1.75 $\pm 0.04$ \\
\\
  & blue-highz  & 2.79  $\pm 0.35$  &  1.53  $\pm 0.14$  &  1.23 $\pm 0.08$ \\
  & red-highz   & 5.88  $\pm 0.21$  &  1.82  $\pm 0.07$  &  2.04 $\pm 0.08$ \\
\\
2 & blue1-lowz  & 3.02  $\pm 0.14$  &  1.60  $\pm 0.10$  &  1.06 $\pm 0.06$ \\
  & blue2-lowz  & 3.76  $\pm 0.16$  &  1.62  $\pm 0.08$  &  1.18 $\pm 0.08$ \\
  & red1-lowz   & 5.46  $\pm 0.32$  &  1.89  $\pm 0.15$  &  1.64 $\pm 0.07$ \\
  & red2-lowz   & 6.82  $\pm 0.39$  &  1.75  $\pm 0.09$  &  1.90 $\pm 0.06$ \\
\\
  & blue1-highz & 2.76  $\pm 0.12$  &  1.58  $\pm 0.11$  &  1.19 $\pm 0.12$ \\
  & blue2-highz & 3.56  $\pm 0.34$  &  1.55  $\pm 0.13$  &  1.45 $\pm 0.16$ \\
  & red1-highz  & 4.92  $\pm 0.55$  &  1.58  $\pm 0.06$  &  1.80 $\pm 0.29$ \\
  & red2-highz  & 7.60  $\pm 0.41$  &  1.91  $\pm 0.10$  &  2.56 $\pm 0.13$ \\
\\
3 & 1-lowz      & 3.11  $\pm 0.69$  &  1.86  $\pm 0.27$  &  0.97 $\pm 0.04$ \\
  & 2-lowz      & 3.27  $\pm 0.20$  &  1.56  $\pm 0.05$  &  1.13 $\pm 0.08$ \\
  & 3-lowz      & 3.55  $\pm 0.18$  &  1.66  $\pm 0.08$  &  1.13 $\pm 0.06$ \\
  & 4-lowz      & 4.88  $\pm 0.35$  &  1.69  $\pm 0.09$  &  1.48 $\pm 0.07$ \\
  & 5-lowz      & 5.92  $\pm 0.17$  &  2.04  $\pm 0.15$  &  1.74 $\pm 0.13$ \\
  & 6-lowz      & 6.67  $\pm 0.42$  &  1.57  $\pm 0.11$  &  1.90 $\pm 0.15$ \\
\\
  & 1-highz     & 2.04  $\pm 0.21$  &  1.41  $\pm 0.06$  &  1.11 $\pm 0.18$ \\
  & 2-highz     & 3.57  $\pm 0.13$  &  1.73  $\pm 0.11$  &  1.35 $\pm 0.05$ \\
  & 3-highz     & 2.54  $\pm 0.45$  &  1.36  $\pm 0.12$  &  1.37 $\pm 0.16$ \\
  & 4-highz     & 3.19  $\pm 0.40$  &  1.66  $\pm 0.17$  &  1.25 $\pm 0.06$ \\
  & 5-highz     & 4.75  $\pm 0.54$  &  1.55  $\pm 0.13$  &  1.80 $\pm 0.17$ \\
  & 6-highz     & 8.29  $\pm 0.38$  &  1.81  $\pm 0.08$  &  2.73 $\pm 0.10$ \\
\enddata
\tablenotetext{*}{These measurements are made on scales of $1 < $ \rp $ < 10$ \mpch.}
\label{tab:wprp}
\end{deluxetable}

\begin{figure*}[th!]
\epsscale{0.75}
\plotone{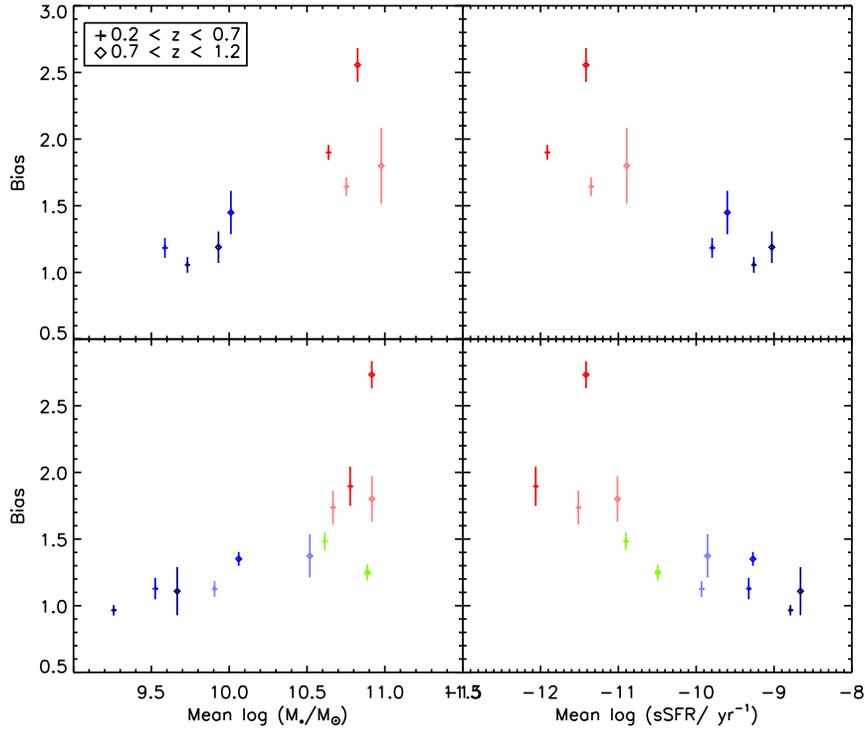}
\caption{\label{fig:bias}
\small The absolute bias on scales $1 < $ \rp $ < 10$ \mpch \ 
of each galaxy sample in runs 2 (``main sequence split'') and 3 (``sSFR cuts'').  The left column shows the bias as a function of \ssfr, and the right column shows the bias as a function of stellar mass. 
 The colors of each sample correspond to the colors shown in Fig.~\ref{fig:samples}. 
}
\end{figure*}

\begin{figure*}[th!]
\epsscale{0.75}
\plotone{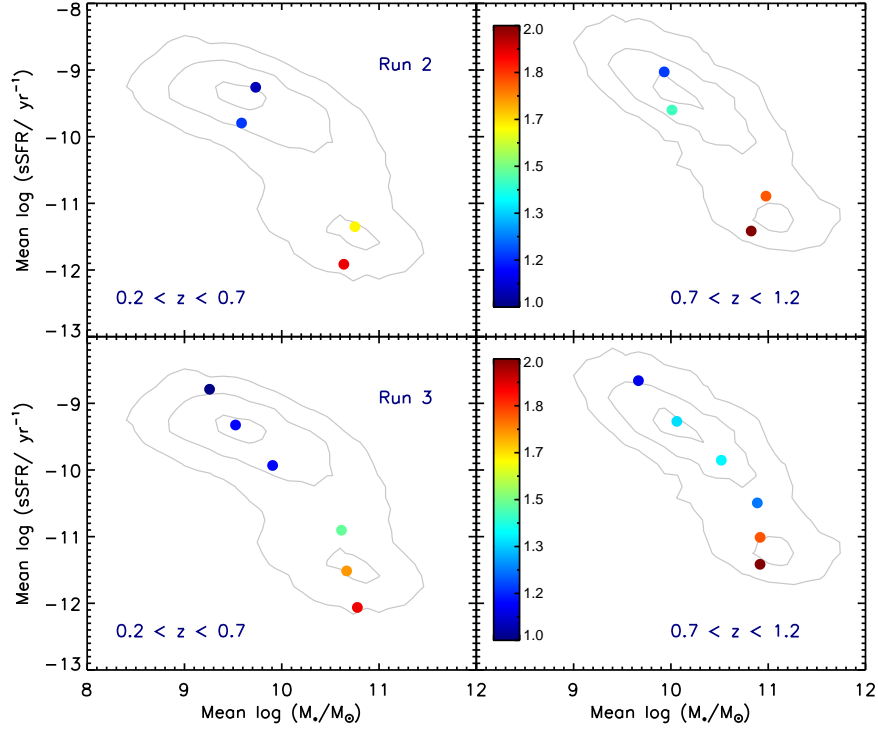}
\caption{\label{fig:jointbias}
\small The absolute bias on scales $1 < $ \rp $ < 10$ \mpch \ 
of each galaxy sample in runs 2 (``main sequence split'') and 3 (``sSFR cuts''), shown here as a joint function of \ssfr and stellar mass.  The color of each point reflects the bias value, as shown in the color bar.  The light grey contours show the full galaxy population in the relevant redshift range.
}
\end{figure*}

\begin{figure*}[th!]
\epsscale{0.75}
\plotone{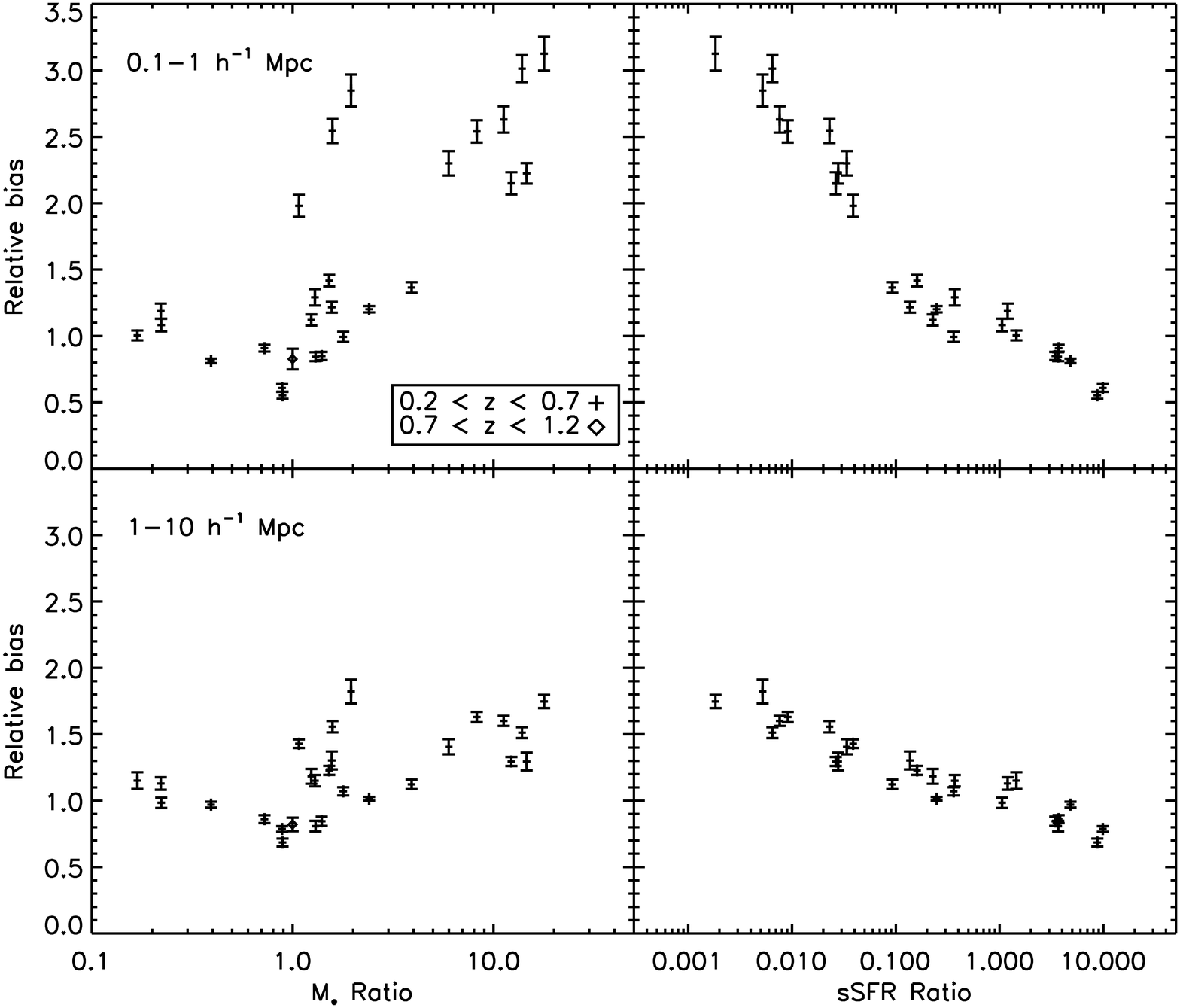}
\caption{\label{fig:relbias}
\small The one-halo (top, $0.1 < $ \rp $ < 1$ \mpch) and two-halo (bottom, $1 < $ \rp $ < 10$ \mpch) relative bias between various galaxy samples, as a function of the stellar mass ratio (left) and \ssfr ratio (right) of the two samples.
Only those relative bias values with an error less than 25\% of the one-halo relative bias are shown, for clarity.  Additional galaxy samples are used here beyond the runs shown earlier in the paper, to help fill in this space.  
It can clearly be seen that the relative bias is more monotonically dependent on 
the \ssfr ratio than the stellar mass ratio.  
}
\end{figure*}

\begin{figure*}[th!]
\epsscale{0.75}
\plotone{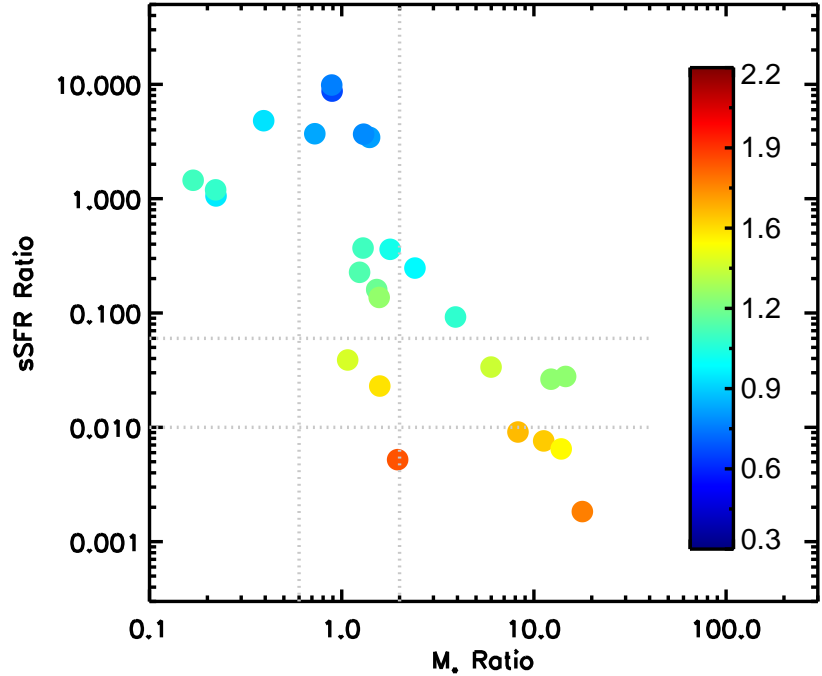}
\caption{\label{fig:jointrelbias}
 \small The two-halo relative bias between various galaxy samples, shown as a joint function of sSFR ratio and stellar mass ratio.  Shown are all relative biases where the fractional error is less than 25\%.  The color of each point reflects the relative bias value, as shown in the color bar.  The dotted line highlight regions of fixed stellar mass or \ssfr ratio where our galaxy samples are able to probe at least an order of magnitude in the ratio of the other parameter (stellar mass or \ssfr).  As seen, at a fixed stellar mass ratio, variations with \ssfr lead to strong differences in the relative bias, while at a fixed \ssfr ratio, variations with stellar mass do not result in substantially different clustering amplitudes. 
}
\end{figure*}

\begin{deluxetable*}{llcccc}[ht!]
\tablecaption{Relative Bias Measurements}
\tablehead{
\colhead{Run}&\colhead{Name}&\colhead{\mass \ Ratio}&\colhead{\ssfr Ratio}&\colhead{Relative Bias}&\colhead{Relative Bias} \\
\colhead{} & \colhead{} & \colhead{} & \colhead{} & \colhead{1-halo}\tablenotemark{*} & \colhead{2-halo} 
}
\startdata
\\
1 & red-lowz/blue-lowz    & 1.1 & 0.039 & 1.98 $\pm 0.08$  & 1.43 $\pm 0.03$  \\
  & red-highz/blue-highz  & 1.1 & 0.059 & 1.56 $\pm 0.14$  & 1.68 $\pm 0.06$ \\
\\
2 & red1-lowz/blue2-lowz  & 15 & 0.028 & 2.22 $\pm 0.08$ & 1.30 $\pm 0.07$  \\
  & red2-lowz/blue2-lowz   & 11 & 0.0076 & 2.63 $\pm 0.10$ & 1.60 $\pm 0.04$   \\
  & blue1-lowz/blue2-lowz & 1.4 & 3.4 & 0.85 $\pm 0.03$   & 0.84 $\pm 0.04$  \\
  & red1-lowz/red2-lowz    & 1.3 & 3.7 & 0.84 $\pm 0.03$    & 0.81 $\pm 0.04$ \\
\\
  & red1-highz/blue2-highz  & 9.3 & 0.051 & 1.99 $\pm 0.19$ & 1.22 $\pm 0.05$ \\
  & red2-highz/blue2-highz  & 6.5 & 0.015 & 2.93 $\pm 0.18$ & 1.79 $\pm 0.06$ \\
  & blue1-highz/blue2-highz & 0.83 & 3.7 & 0.82 $\pm 0.05$  & 0.80 $\pm 0.04$  \\
  & red1-highz/red2-highz   & 1.4 & 3.3 & 0.80 $\pm 0.09$  & 0.68 $\pm 0.03$ \\
\\
3 & 1-lowz/2-lowz           & 0.54 & 3.4 & 1.06 $\pm 0.11$ & 0.84 $\pm 0.05$ \\
  & 3-lowz/2-lowz           & 2.4 & 0.25 & 1.20 $\pm 0.02$ & 1.01 $\pm 0.01$ \\
  & 4-lowz/2-lowz           & 12 & 0.026 & 2.15 $\pm 0.08$ & 1.30 $\pm 0.03$ \\
  & 5-lowz/2-lowz           & 14 & 0.0065 & 3.01 $\pm 0.10$ & 1.51 $\pm 0.04$\\
  & 6-lowz/2-lowz           & 18 & 0.0018 & 3.13 $\pm 0.13$ & 1.75 $\pm 0.05$\\
\\
  & 1-highz/2-highz         & 0.40 & 4.1 & 0.92 $\pm 0.11$   & 0.81 $\pm 0.04$ \\
  & 3-highz/2-highz         & 2.9 & 0.26 & 1.41 $\pm 0.11$ & 1.02 $\pm 0.03$ \\
  & 4-highz/2-highz         & 6.7 & 0.060 & 1.80 $\pm 0.25$  & 0.92 $\pm 0.06$ \\
  & 5-highz/2-highz         & 7.2 & 0.018 & 2.18 $\pm 0.23$  & 1.33 $\pm 0.09$ \\
  & 6-highz/2-highz         & 7.2 & 0.0072 & 3.97 $\pm 0.32$ & 2.06 $\pm 0.08$ \\
\\
4 & 1-lowz/3-lowz          & 0.22 & 1.2 & 0.79 $\pm 0.07$   & 1.05 $\pm 0.08$ \\
  & 2-lowz/4-lowz           & 0.17 & 1.4 & 1.00 $\pm 0.04$   & 1.15 $\pm 0.06$  \\
  & 4-lowz/7-lowz           & 0.17 & 1.6 & 0.86 $\pm 0.06$   & 1.04 $\pm 0.03$  \\
  & 5-lowz/8-lowz           & 0.22 & 1.1 & 1.08 $\pm 0.05$   & 0.98 $\pm 0.04$  \\
  & 6-lowz/9-lowz           & 0.22 & 1.2 & 1.19 $\pm 0.06$ &  1.13 $\pm 0.05$  \\
  & 1-lowz/2-lowz           & 0.87 & 3.4 & 0.85 $\pm 0.04$ & 0.82 $\pm 0.04$  \\
  & 3-lowz/4-lowz           & 0.67 & 4.2 & 1.17 $\pm 0.11$  & 0.91 $\pm 0.04$  \\
  & 5-lowz/6-lowz           & 0.89 & 8.7 & 0.55 $\pm 0.03$ & 0.69 $\pm 0.03$  \\
  & 7-lowz/8-lowz           & 0.85 & 6.7 & 0.75 $\pm 0.05$ & 0.77 $\pm 0.03$  \\
  & 8-lowz/9-lowz           & 0.89 & 9.7 & 0.61 $\pm 0.03$ & 0.79 $\pm 0.02$  \\
\\                                 
  & 1-high/2-high           & 0.27 & 1.5 & 0.59 $\pm 0.08$ &  0.98 $\pm 0.07$ \\
  & 3-high/5-high           & 0.22 & 1.8 & 0.80 $\pm 0.07$ & 1.08 $\pm 0.04$  \\
  & 4-high/6-high           & 0.26 & 1.2 & 0.84 $\pm 0.14$ & 0.56 $\pm 0.09$  \\
  & 3-high/4-high           & 0.61 & 12  & 1.01 $\pm 0.88$ &  1.66 $\pm 0.55$ \\
  & 4-high/5-high           & 0.36 & 0.15 & 0.96 $\pm 0.19$ & 0.69 $\pm 0.13$ \\
  & 5-high/6-high           & 0.72 & 8.2 & 0.83 $\pm 0.08$  & 0.82 $\pm 0.05$  \\

\enddata
\tablenotetext{*}{The ``1-halo'' relative bias measurements are on scales 
$0.1 < $ \rp $ < 1$ \mpch \ and the ``2-halo'' measurements are on scales $1 < $ \rp $ < 10$ \mpch.}
\label{tab:relbias}
\end{deluxetable*}

In Fig.~\ref{fig:bias} we show the dependence of the large-scale bias, measured on scales $1 < $ \rp $ < 10$ \mpch, on stellar mass (left) and \ssfr (right).  We show results for galaxy samples in runs 2 (``main sequence split'', top row) and 3 (``sSFR cuts'', bottom row) in this figure.  While there is a general increase in the bias with increasing stellar mass, there are specific samples where the bias is not necessarily higher at higher stellar mass.  As showed in the right panels, however, there is a clear steady increase in the bias with decreasing \ssfr, such that the trend is monotonic with \ssfr.  We show results for both redshift ranges overlaid, and note that within the errors, at a given stellar mass the bias does not show any redshift dependence, but at a given \ssfr the bias at higher redshift is higher than the bias at lower redshift (i.e., the diamonds lie above the crosses in the left column of Fig.~\ref{fig:bias}).  

As the clustering of dark matter particles increases with time, the bias between a galaxy population and dark matter particles generally {\it decreases} over time.  Therefore one would assume that the absolute bias at a given galaxy property should be higher at higher redshift.  However, both the stellar mass and \ssfr change with time for individual galaxies, and the \ssfr of a galaxy changes {\it more} between $z\sim0.9$ and $z\sim0.5$ than the stellar mass at \logmass=10.5 (corresponding to the mean stellar mass probed here) \citep[i.e.,][]{Moustakas13, Madau14}.
We return to this point in the discussion section below.

In Fig.~\ref{fig:jointbias} we show with colored points the absolute bias of various galaxy samples from runs 2 and 3 as a function of stellar mass on the x-axis and \ssfr on the y-axis.  The light grey contours show the full galaxy population in our sample in the redshift range of interest.  This figure allows one to clearly see how the bias is changing as a function of \ssfr \ {\it at a given stellar mass}, both within the star-forming and quiescent populations.  Generally, we find that the bias increases towards the lower right of this figure, at higher stellar mass and lower \ssfr.  We return in the discussion section below to how galaxies likely evolve in this plane.

\subsection{Relative Bias Between Galaxy Samples} \label{sec:relbias}

We also quantify how the {\it relative} bias between two galaxy samples depends on both stellar mass and \ssfr.  The relative bias between two galaxy samples can have smaller errors than a direct comparison of the absolute bias values, as to first order cosmic variance effects will cancel when comparing the clustering of two galaxy samples in the same volume.  We may therefore be able to obtain more significant dependences on how the relative bias depends on stellar mass and \ssfr than quantifying only the absolute bias dependence on these parameters.

The relative bias between various galaxy subsamples from runs 2, 3, and 4 are 
listed in Table~\ref{tab:relbias}. We quantify the relative bias on two scales: the ``one-halo'' scale of $0.1 < $ \rp $ < 1$ \mpch \ and the ``two-halo'' scale of $1 < $ \rp $ < 10$ \mpch.  Here again, as with the absolute bias, we find that star-forming galaxies above the main sequence are less clustered than star-forming galaxies below the main sequence (in run 2, ``main sequence split'', in both redshift ranges there is a $3-5\sigma$ difference on both small and larger scales).  We also find that among the quiescent galaxy population, those galaxies with a higher SFR at a given stellar mass are less clustered (in run 2 at lower redshift there is a 5$\sigma$ difference on small and large scales, while at higher redshift there is an 11$\sigma$ difference on large scales).  Significant differences within the star-forming and quiescent populations can also be seen in the results for run 3 (``sSFR cuts''), using finer bins in \ssfr.

 We also list the stellar mass and \ssfr ratio between the two relevant galaxy samples in Table~\ref{tab:relbias}.  These are defined as \mass$_1$/\mass$_2$ and sSFR$_1$/sSFR$_2$ where 1 and 2 correspond to the galaxy samples of interest, where the relative bias is the square root of the ratio of \wp of sample 1 to \wp of sample 2.  A stellar mass or \ssfr ratio near unity reflects that the two galaxy samples of interest have similar stellar mass or \ssfr, while ratios much larger than unity reflects that sample 1 has a much higher stellar mass or \ssfr (i.e., is more highly star-forming) than sample 2.  Values of these ratios that are less than unity reflect that sample 2 has a higher stellar mass or \ssfr than sample 1.

The relative bias as a function of stellar mass and \ssfr ratio are shown in Fig.~\ref{fig:relbias}.  The relative bias on small, ``one-halo'' scales is shown on the top, while the relative bias of the same samples on larger, ``two-halo'' scales is shown on the bottom.  
We show results from runs 1, 2, 3 and 4 as well as some additional galaxy subsamples made to help create a more even distribution in stellar mass and \ssfr ratios (i.e., with stellar mass ratio near unity and \ssfr ratio between 0.1 to 1.0).  These additional samples are very similar to those in runs 1 (star-forming/quiescent split) and 2 (main sequence split), we simply further divide the star-forming and quiescent populations into more bins, using either a simple cut in \ssfr or using the tilt of the main sequence as in run 2.
Instead of plotting all relative bias results, we show only those where the ``one-halo'' error is less than 5\% of the relative bias; thus only high signal-to-noise results are shown.  
We show results from both redshift ranges used here and find that the {\it relative} bias between galaxy samples does not evolve strongly with redshift within the range probed here, as expected. 

We find that the relative bias is a very smooth function of the \ssfr ratio, declining steadily as the \ssfr ratio increases, on both small and large scales.  However, the relative bias is {\it not} as smooth of a function of the stellar mass ratio; at a stellar mass ratio of $\sim$0.7--2, there is a wide range of relative biases, between $\sim$0.7--2.  
We also note that all of the high relative bias values ($>2$ on small scales)
have very low \ssfr ratios ($<0.05$), while they have a range of stellar mass ratios (1--20).  
We also find that the relative bias values are more extreme on small scales than on large scales.  While the same trends are seen on both one- and two-halo scales, the trend is stronger on small scales.  

In order to more clearly understand the dependence of the relative bias on stellar mass and \ssfr, we show in Fig.~\ref{fig:jointrelbias} the {\it joint} dependence of the two-halo relative bias on the stellar mass and \ssfr ratio (the one-halo relative bias shows the same trends in this space.)  We highlight with dotted lines two regions of the diagram where there are multiple samples with a relatively narrow range in one ratio and a wider range in the other ratio. For example, at a stellar mass ratio 
near unity there are many points spanning a \ssfr ratio of $\sim$0.01--10.  As can be seen in the figure, the relative bias of these points --- at a fixed stellar mass ratio --- varies substantially, and monotonically, as the \ssfr ratio varies.  However, at a fixed \ssfr ratio near $\sim$0.03, for points where the stellar mass ratio varies from $\sim$1--10, there is very little change in the relative bias.  This clearly shows that the relative bias depends strongly on the \ssfr ratio, even at a given stellar mass ratio, while the same is not true of the stellar mass ratio at given \ssfr ratio.  
Therefore, the dependence of galaxy clustering on \ssfr is {\it stronger} at a fixed stellar mass than the dependence on stellar mass at a fixed \ssfr.

\section{Discussion} \label{sec:discussion}
In this section we compare our results to the relevant literature and 
discuss how our findings place constraints on how galaxies evolve in 
the \ssfr-\mass plane.  We also discuss how these results impact our 
understanding of the mapping between galaxies and dark matter halos, 
including expanding the halo model of galaxy evolution to explicitly 
include \ssfr.  

\begin{figure}[t!]
\epsscale{1.2}
\plotone{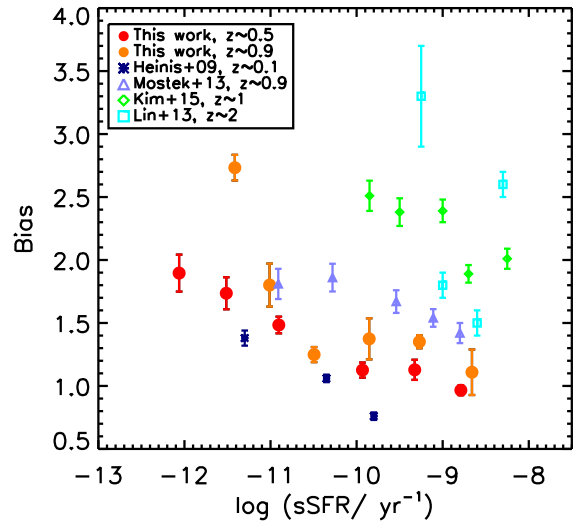}
\caption{\label{fig:literature}
\small The two-halo absolute bias of galaxies as a function of sSFR,
for our results compared to other results in the literature at $z\sim1-2$ 
\citep{Lin12,Mostek13,Kim15}. 
}
\end{figure}

\subsection{Comparison with Literature}

There are many measurements in the literature of the relative bias of 
quiescent to star-forming galaxies, either at a given magnitude or stellar 
mass, where these two galaxy populations are defined by color, sSFR,
or spectral type.  Here, at a stellar mass of \mass$\sim$10.5, we find a 
relative bias of 2.2 ($\pm$0.3) on one-halo scales and 1.5 ($\pm$0.1) on 
two-halo scales at $z=$0.5 and a relative bias of 1.9 ($\pm$0.6) on 
one-halo scales and 1.7 ($\pm0.2$) on two-halo scales 
at $z=$0.9.  Similar values are typically found by others, with values 
on large scales of $\sim1.3-1.5$ \citep[e.g.,][]{Madgwick03, Meneux06, 
Coil08, delaTorre11, Hearin14}. The relative bias of quiescent to star-forming
galaxies can be a strong function of scale, however, and may be a 
function of stellar mass as well, so care should be taken when comparing 
results from different surveys.

There are also relevant results in the literature from weak lensing
studies, which directly measure halo masses around selected galaxy samples.
Several studies using galaxy color have found that at a given stellar
mass, red galaxies have larger halo masses than blue galaxies 
\citep{Velander14, Rodriguez-Puebla15, Zu16}, similar to the trends
found here and in other clustering studies.  \citet{Mandelbaum16} use 
locally brightest galaxies in SDSS as a proxy for selecting central
galaxies, in order to quantify how halo mass depends on stellar mass and color 
for central galaxies.  They find that for \logmass $>$ 10.7, red central
galaxies are in halos that are at least twice as massive as those of
blue central galaxies, again qualitatively consistent with our
findings, although we do not attempt to isolate central galaxies.

The most relevant papers to compare our results to here are those that study 
the dependence of clustering on multiple bins in \ssfr. We present the 
bias reported as a function of \ssfr for the relevant papers in 
Fig.~\ref{fig:literature}.  
Our results are somewhat higher than those of \citet{Heinis09}, but 
given that their sample is at $z\sim0.1$ this is to be expected, as 
the bias decreases with cosmic time. 
Our results agree fairly well with those of 
\citet{Mostek13}, using the full DEEP2 survey at $z\sim0.9$, and are 
much lower than \citet{Kim15}, also at $z\sim1$.  We also compare with results 
from \citet{Lin12} at $z\sim2$.  
The bias should decrease from $z\sim2$ to $z\sim1$
due to the increased clustering of dark matter particles at lower redshift;
the change in the growth factor over this redshift interval is 
$\sim$45\%.  
Given this, it is not unexpected that our results are lower than those of 
\citet{Lin12}; however we note that two of their four data points have very 
high bias values ($>$2.5) and they do not find a monotonic trend in the bias 
with \ssfr. The \citet{Lin12} and \citet{Kim15} results are derived 
from angular clustering measurements, which may impact their robustness.
Generally, however, these results show that at a given redshift the bias 
decreases with increasing \ssfr.  

When comparing the clustering of galaxies as a function of \ssfr \ at 
different redshifts, of course, the overall evolution in the normalization 
of the star-forming main sequence should be taken into account 
\citep[e.g.,][]{Karim11,Whitaker14,Speagle14}.  However, this evolution 
by itself can not account for the \ssfr dependence in clustering that is observed 
at a given redshift.  The scatter in the main sequence does not evolve
substantially with cosmic time \citep{Speagle14}, such that essentially
the \ssfr of the bulk of star-forming galaxies is decreasing with time 
since $z\sim2$.  The fact that at a given epoch there is a correlation 
between clustering amplitude and \ssfr \ therefore implies that galaxies 
evolve from having {\it relatively} high \ssfr (with respect to the main 
sequence at that redshift) to relatively low \ssfr.  This is discussed more
in the next section below.

We do not compare our results to stellar mass-dependent clustering
results in the literature, in part because the PRIMUS stellar
mass-dependent clustering is presented in \citet{Skibba15} and also
because here we have not neccessarily included samples that are complete for
all sSFRs at a given stellar mass (other than run 1,
``star-forming/quiescent split'').  
We note, however, that not all
papers that investigate the stellar mass-dependent clustering of
galaxies do use samples that are complete to all galaxy types (i.e.,
all sSFRs).  Furthermore, because stellar mass and sSFR are not fully
independent quantities for galaxies --- there are correlations between
them --- results on the stellar mass-dependence of clustering are likely
impacted by differences in the sSFR of the samples.  As pointed out
by \citet{Coil08} and others, the luminosity-dependence of the clustering of all
galaxies is stronger than the luminosity-dependence present in either
the star-forming or quiescent populations alone.  Essentially, part of
the overall luminosity-dependence that is observed is due to the
changing fraction of quiescent galaxies (which are more clustered) as
a function of luminosity.  The same holds for stellar mass and sSFR,
in that the most massive galaxies have lower sSFRs, on average, at
$z\lesssim2$.  Therefore, much of what has been interpreted as
differences in galaxy clustering due to stellar mass may be driven in
part by differences in sSFR.

\subsection{Evolution of Galaxies in sSFR-\mass \ Plane}

One of the main findings of this paper is that at intermediate redshift 
the large-scale galaxy clustering amplitude smoothly increases 
across the \ssfr-\mass \ plane, from
lower mass galaxies that are forming stars at a high rate 
(low \mass, high \ssfr, the upper left in the lower panels of 
Fig.~\ref{fig:jointbias}) 
to higher mass galaxies that are forming stars at a very low rate 
(high \mass, low \ssfr, the lower right of this figure).  
As the clustering of a given coeval galaxy population can generally only 
increase over time, this implies that galaxies {\it evolve} across 
the \ssfr-\mass \ plane from the upper left to the lower right.  
A similar conclusion is reached by 
\citet{Kim15} for central galaxies, using halo occupation distribution 
modeling of their stellar mass and \ssfr-dependent clustering results at 
$z\sim1$.    

This implies, interestingly, that star-forming galaxies do not simply
evolve solely {\it along} the main sequence of star formation, increasing their SFR
as their stellar mass increases.  This is shown 
by the fact that at a given stellar mass, the clustering of star-forming 
galaxies above
the main sequence is lower than that of star-forming galaxies below the main 
sequence (see also Mostek et al. 2013).  
This should not perhaps be surprising given the known differences in other
galaxy physical parameters above and below the main sequence, such as 
SFR surface densities, sizes, dust properties, and Sersic index 
\citep{Schiminovich07, Elbaz11, Wuyts11}.  These different physical properties 
are often interpreted as being due to galaxies above the main sequence 
experiencing merger events, but our results here do not support this 
interpretation.  We do not find a rise in the clustering amplitue on small
scales for star forming galaxies above the main sequence compared to below
the main sequence, and given the differences in large-scale clustering 
amplitude between these populations, it is clear that galaxies must begin
their lives above the main sequence and evolve across it.  It is therefore
likely that galaxies do not solely move along the ridge of the main sequence 
as they grow.

We also conclude that the higher clustering amplitude seen in other studies 
for galaxies with higher SFR is not simply due to the fact that they have 
higher stellar masses (i.e., due to the star-forming main sequence).  If the 
increase in clustering was entirely driven by differences in the stellar mass, 
that would imply that there should be no difference in the clustering amplitude 
{\it along} the main sequence of star formation.  However, as shown by the 
results here for run 3 (``sSFR cuts'') at $0.7 < z < 1.2$, we find a significant increase 
in the bias value at the more massive end of the main sequence than at lower 
stellar mass on the sequence.  Therefore, there must be an additional
clustering dependence on SFR or sSFR beyond the stellar mass
dependence.  Indeed, that is what we find when we compare the relative
bias of galaxies as a function of both the stellar mass 
and \ssfr ratio of the relevant galaxy samples.

\subsection{Connecting Galaxies and Dark Matter Halos}

We have shown that at intermediate redshift galaxy clustering
correlates more strongly with \ssfr than with stellar mass.  
A similar conclusion was also drawn by \citet{Heinis09} with a smaller 
sample of \ssfr bins. This
conclusion is also similar to that of \citet{Coil08}, who found that
at $z\sim1$ the
dependence of clustering on color is much stronger than with
luminosity, given that color is highly correlated with sSFR and luminosity
correlates with stellar mass.  

We find that the stellar mass of a galaxy does correlate with
clustering amplitude and therefore halo mass, but much of this dependence 
appears to be driven by differences in \ssfr.  There is a correlation in the
galaxy population between stellar mass and \ssfr, and while higher
stellar mass galaxies are more clustered \citep[e.g.,][]{Skibba14}, we 
do not find this to be true at a given \ssfr.

The halo model of galaxy evolution essentially posits that the dark matter
halo mass that a galaxy resides in determines all of the galaxy's properties
\citep[e.g.,][]{Peacock00, Seljak00}.  Our results would seem to counter 
that, in that clearly a 
given halo mass can correspond to a range of stellar masses for a galaxy.  
It is therefore not straightforward to predict the stellar mass of a galaxy, 
simply from knowing the halo mass that it resides in, as \ssfr is another key 
parameter.
Indeed, age-matching models \citep{Hearin13} predict that at a given 
stellar mass, star-forming galaxies are less clustered than quiescent
galaxies, as we find here.  This would seem to imply that our results 
favor age-matching-type models over strict halo models of galaxy evolution.  

However, our results can not rule out the halo model, as at a given stellar 
mass, satellite galaxies are found to reside in somewhat more massive 
halos than 
central galaxies \citep{Watson13} and satellite galaxies of a given stellar 
mass (in more massive halos) are more likely to be quiescent than central
galaxies in lower mass halos \citep{Wetzel12}.  
This can lead to a higher clustering
amplitude for quiescent galaxies compared to star-forming galaxies at a 
given stellar mass, without invoking assembly bias, which posits that 
another property of the halo besides mass is relevant in determining 
galaxy properties.  In other words, both star-forming and quiescent 
central galaxies of the same stellar mass could have the same clustering 
amplitude, while comparing the clustering of all galaxies (including 
satellites) of that stellar mass, quiescent galaxies would be more 
clustered, as they include more satellites in higher mass halos.
Our results therefore do not neccessarily imply assembly bias; a detailed
comparison with halo models and models that include assembly bias is 
required to make that claim.  However, \citet{Berti16} measure the galaxy 
conformity signal in PRIMUS, essentially through a cross-correlation of 
isolated, massive galaxies with lower mass, star-forming galaxies, and 
find a signal that likely does reflect assembly bias at these redshifts.

We note that much of the stellar mass dependence that previous papers have 
found in galaxy clustering may be influenced by differences in the 
\ssfr of the galaxy samples used, given the correlation between stellar mass 
and \ssfr within the full galaxy population (and within the star-forming 
and quiescent populations separately).  
We find that 
at a given stellar mass, star-forming galaxies are significantly less 
clustered than quiescent galaxies, which implies that halo mass
must depend jointly on stellar mass and \ssfr.  The stellar mass to halo 
mass relation, then, likely needs to be expanded to account for \ssfr.  
This would
essentially involve shifting the stellar mass to halo mass relation to 
higher or lower halo masses depending on the \ssfr of the galaxy in question.
The scatter that 
has been quantified in the stellar mass to halo mass relation 
\citep[e.g.,][]{More11,Moster13,Behroozi13} may be due in part to \ssfr.
The \ssfr of a galaxy appears to be in fact {\it more} correlated with halo
mass than stellar mass correlates with halo mass.  While it is likely that 
much of this reflects that the red fraction of satellite galaxies
increases with halo mass (even at a given stellar mass, e.g., Prescott
et al. 2011)\nocite{Prescott11}, it 
also seems likely that even for central galaxies there is a dependence on \ssfr
at a given halo mass.  Indeed, below the break in the stellar mass function, 
halo mass does not strongly correlate with stellar mass, though our
results suggest that it may correlate with \ssfr.

Finally, we note that the relative bias results presented here as a function 
of the joint dependence on the stellar mass ratio and \ssfr ratio
provide  
very strong constraints for theoretical models of galaxy evolution.
They are also a new way of using the data to measure the dependence of galaxy
clustering on these parameters.  This new measurement of the joint dependence of
the {\it relative} bias on ratios of galaxy properties should help 
differentiate between competing theoretical models of galaxy evolution.

\section{Conclusions} \label{sec:conclusions}
In this paper we have used the PRIMUS and DEEP2 galaxy redshift surveys to 
study the joint dependence of galaxy clustering properties on stellar mass 
and \ssfr.  We utilize a full sample of over 100,000 spectroscopic redshifts 
to divide our sample into two redshift ranges, $0.2 < z < 0.7$ and 
$0.7 < z < 1.2$ and use SED fits to estimate the galaxy stellar mass and \ssfr.
We divide the full galaxy population not only into star-forming and quiescent 
samples, but we subdivide each of these populations according to \ssfr or 
distance from the main sequence of star formation, to study the dependence 
of the clustering amplitude in relatively fine bins in \ssfr.  
We measure both the absolute bias of galaxy samples with respect to dark matter 
and also the relative bias between galaxy samples, as a joint function of
the ratio of the stellar masses and sSFRs of the galaxy samples.

Our main conclusions are:

\begin{itemize}

\item Galaxy clustering depends just as strongly on sSFR as on stellar mass, within 
the stellar mass range probed here.  Our results imply that the stellar mass to halo mass relation may depend on sSFR as well.

\item Within the star-forming population at a given stellar mass, galaxies with a high sSFR that lie above the main sequence are less clustered than galaxies with a relatively low sSFR below the main sequence. This is also true within the quiescent population, in that galaxies with a higher sSFR are less clustered than galaxies with a lower sSFR, at a given stellar mass.  This constrains the evolutionary path of galaxies in the sSFR-stellar mass plane, indicating that they likely evolve from high sSFR and lower stellar mass to low sSFR and higher stellar mass.  In particular, galaxies likely evolve {\it across} the main sequence of star formation, not only along it, before becoming quiescent.  Within the quiescent population, galaxies with higher sSFR are likely also younger, on average, than those with lower sSFR.

\item We present new measurements of the relative bias of galaxies as a {\it joint} function of the stellar mass ratio and sSFR ratio of galaxy samples, showing that at a given stellar mass ratio there is a strong dependence of clustering amplitude on the sSFR ratio.  The reverse is not true, however; at a given sSFR ratio there does not appear to be a strong dependence of the clustering amplitude on the stellar mass ratio.  This shows that while galaxy clustering depends on stellar mass, it does not depend on 
stellar mass {\it at a given sSFR}, within the range of stellar mass and sSFR probed here.

\end{itemize} 

These results are strongly constraining for theoretical models of galaxy
evolution, both for age-matching and other empirically-based 
methods \citep[e.g.,][]{Behroozi13}, as well as semi-analytic models. 
It would clearly be beneficial to perform similar investigations at both 
lower and higher redshift. Such measurements, undertaken across a range of
redshifts and cosmic time, would be extremely constraining for theoretical 
models of galaxy evolution and the galaxy-halo connection.

\acknowledgements

We thank Andrew Hearin and Peter Behroozi for useful discussions and
the referee for detailed comments.
Funding for PRIMUS has been provided by NSF grants AST-0607701, 0908246,
0908442, 0908354, and NASA grant 08-ADP08-0019. ALC acknowledges support from
NSF CAREER award AST-1055081.
 We use
data from the DEEP2 survey, which was supported by NSF AST grants AST00-71048,
AST00-71198, AST05-07428, AST05-07483, AST08-07630, AST08-08133. This study
makes use of data from AEGIS Survey and in particular uses data from $GALEX$,
$Spitzer$, CFHT, and the Keck and Magellan telescopes.


\begin{thebibliography}{}
\expandafter\ifx\csname natexlab\endcsname\relax\def\natexlab#1{#1}\fi

\bibitem[{{Alimi} {et~al.}(1988){Alimi}, {Valls-Gabaud}, \&
  {Blanchard}}]{Alimi88}
{Alimi}, J.-M., {Valls-Gabaud}, D., \& {Blanchard}, A. 1988, \aap, 206, L11

\bibitem[{{Baldry} {et~al.}(2004)}]{Baldry04}
{Baldry}, I.~K., {et~al.} 2004, \apj, 600, 681

\bibitem[{{Behroozi} {et~al.}(2010){Behroozi}, {Conroy}, \&
  {Wechsler}}]{Behroozi10}
{Behroozi}, P.~S., {Conroy}, C., \& {Wechsler}, R.~H. 2010, \apj, 717, 379

\bibitem[{{Behroozi} {et~al.}(2013){Behroozi}, {Wechsler}, \&
  {Conroy}}]{Behroozi13}
{Behroozi}, P.~S., {Wechsler}, R.~H., \& {Conroy}, C. 2013, \apj, 770, 57

\bibitem[{{Benoist} {et~al.}(1996){Benoist}, {Maurogordato}, {da Costa},
  {Cappi}, \& {Schaeffer}}]{Benoist96}
{Benoist}, C., {Maurogordato}, S., {da Costa}, L.~N., {Cappi}, A., \&
  {Schaeffer}, R. 1996, \apj, 472, 452

\bibitem[{{Berlind} {et~al.}(2005)}]{Berlind05}
{Berlind}, A.~A., {et~al.} 2005, ApJ, 629, 625

\bibitem[{{Berti} {et~al.}(2016)}]{Berti16}
{Berti}, A., {et~al.} 2016, ApJ, submitted, arXiv 1409.7071

\bibitem[{{Bigelow} \& {Dressler}(2003)}]{Bigelow03}
{Bigelow}, B.~C., \& {Dressler}, A.~M. 2003, in Society of Photo-Optical
  Instrumentation Engineers (SPIE) Conference Series, Vol. 4841, Society of
  Photo-Optical Instrumentation Engineers (SPIE) Conference Series, ed.
  M.~{Iye} \& A.~F.~M. {Moorwood}, 1727

\bibitem[{{Blanton} \& {Roweis}(2007)}]{Blanton07}
{Blanton}, M.~R., \& {Roweis}, S. 2007, \aj, 133, 734

\bibitem[{{Bruzual} \& {Charlot}(2003)}]{Bruzual03}
{Bruzual}, G., \& {Charlot}, S. 2003, \mnras, 344, 1000

\bibitem[{{Chabrier}(2003)}]{Chabrier03}
{Chabrier}, G. 2003, \pasp, 115, 763

\bibitem[{{Charlot} \& {Fall}(2000)}]{Charlot00}
{Charlot}, S., \& {Fall}, S.~M. 2000, \apj, 539, 718

\bibitem[{{Coil} {et~al.}(2004{\natexlab{a}}){Coil}, {Newman}, {Kaiser},
  {Davis}, {Ma}, {Kocevski}, \& {Koo}}]{Coil04a}
{Coil}, A.~L., {Newman}, J.~A., {Kaiser}, N., {et~al.} 2004{\natexlab{a}},
  \apj, 617, 765

\bibitem[{{Coil} {et~al.}(2004{\natexlab{b}})}]{Coil04}
{Coil}, A.~L., {et~al.} 2004{\natexlab{b}}, \apj, 609, 525

\bibitem[{{Coil} {et~al.}(2006)}]{Coil06}
---. 2006, \apj, 644, 671

\bibitem[{{Coil} {et~al.}(2008)}]{Coil08}
---. 2008, \apj, 672, 153

\bibitem[{{Coil} {et~al.}(2011)}]{Coil11}
---. 2011, \apj, 741, 8

\bibitem[{{Cool} {et~al.}(2013)}]{Cool13}
{Cool}, R.~J., {et~al.} 2013, \apj, 767, 118

\bibitem[{{Davis} \& {Peebles}(1983)}]{Davis83}
{Davis}, M., \& {Peebles}, P.~J.~E. 1983, \apj, 267, 465

\bibitem[{{de la Torre} {et~al.}(2011)}]{delaTorre11}
{de la Torre}, S., {et~al.} 2011, \mnras, 412, 825

\bibitem[{{Dolley} {et~al.}(2014){Dolley}, {Brown}, {Weiner}, {Brodwin},
  {Kochanek}, {Pimbblet}, {Palamara}, {Jannuzi}, {Dey}, {Atlee}, \&
  {Beare}}]{Dolley14}
{Dolley}, T., {Brown}, M.~J.~I., {Weiner}, B.~J., {et~al.} 2014, \apj, 797, 125

\bibitem[{{Durkalec} {et~al.}(2015)}]{Durkalec15}
{Durkalec}, A., {et~al.} 2015, \aap, 576, L7

\bibitem[{{Elbaz} {et~al.}(2011)}]{Elbaz11}
{Elbaz}, D., {et~al.} 2011, \aap, 533, A119

\bibitem[{{Faber} {et~al.}(2003)}]{Faber03}
{Faber}, S.~M., {et~al.} 2003, in Society of Photo-Optical Instrumentation
  Engineers (SPIE) Conference Series, Vol. 4841, Society of Photo-Optical
  Instrumentation Engineers (SPIE) Conference Series, ed. M.~{Iye} \& A.~F.~M.
  {Moorwood}, 1657

\bibitem[{{Hearin} \& {Watson}(2013)}]{Hearin13}
{Hearin}, A.~P., \& {Watson}, D.~F. 2013, \mnras, 435, 1313

\bibitem[{{Hearin} {et~al.}(2014){Hearin}, {Watson}, {Becker}, {Reyes},
  {Berlind}, \& {Zentner}}]{Hearin14}
{Hearin}, A.~P., {Watson}, D.~F., {Becker}, M.~R., {et~al.} 2014, \mnras, 444,
  729

\bibitem[{{Heinis} {et~al.}(2009)}]{Heinis09}
{Heinis}, S., {et~al.} 2009, \apj, 698, 1838

\bibitem[{{Hogg} {et~al.}(2003)}]{Hogg03}
{Hogg}, D.~W., {et~al.} 2003, \apjl, 585, L5

\bibitem[{{Karim} {et~al.}(2011)}]{Karim11}
{Karim}, A., {et~al.} 2011, \apj, 730, 61

\bibitem[{{Kim} {et~al.}(2015){Kim}, {Im}, {Lee}, {Edge}, {Wake}, {Merson}, \&
  {Jeon}}]{Kim15}
{Kim}, J.-W., {Im}, M., {Lee}, S.-K., {et~al.} 2015, \apj, 806, 189

\bibitem[{{Landy} \& {Szalay}(1993)}]{Landy93}
{Landy}, S.~D., \& {Szalay}, A.~S. 1993, \apj, 412, 64

\bibitem[{{Leauthaud} {et~al.}(2011){Leauthaud}, {Tinker}, {Behroozi}, {Busha},
  \& {Wechsler}}]{Leauthaud11}
{Leauthaud}, A., {Tinker}, J., {Behroozi}, P.~S., {Busha}, M.~T., \&
  {Wechsler}, R.~H. 2011, \apj, 738, 45

\bibitem[{{Leauthaud} {et~al.}(2012)}]{Leauthaud12}
{Leauthaud}, A., {et~al.} 2012, \apj, 744, 159

\bibitem[{{Li} {et~al.}(2008){Li}, {Kauffmann}, {Heckman}, {Jing}, \&
  {White}}]{Li08}
{Li}, C., {Kauffmann}, G., {Heckman}, T.~M., {Jing}, Y.~P., \& {White},
  S.~D.~M. 2008, \mnras, 385, 1903

\bibitem[{{Li} {et~al.}(2006)}]{Li06}
{Li}, C., {et~al.} 2006, \mnras, 368, 21

\bibitem[{{Lin} {et~al.}(2012)}]{Lin12}
{Lin}, L., {et~al.} 2012, \apj, 756, 71

\bibitem[{{Lonsdale} {et~al.}(2003)}]{Lonsdale03}
{Lonsdale}, C.~J., {et~al.} 2003, \pasp, 115, 897

\bibitem[{{Loveday} {et~al.}(1995){Loveday}, {Maddox}, {Efstathiou}, \&
  {Peterson}}]{Loveday95}
{Loveday}, J., {Maddox}, S.~J., {Efstathiou}, G., \& {Peterson}, B.~A. 1995,
  \apj, 442, 457

\bibitem[{{Madau} \& {Dickinson}(2014)}]{Madau14}
{Madau}, P., \& {Dickinson}, M. 2014, \araa, 52, 415

\bibitem[{{Madgwick} {et~al.}(2003)}]{Madgwick03}
{Madgwick}, D.~S., {et~al.} 2003, \mnras, 344, 847

\bibitem[{{Mandelbaum} {et~al.}(2016){Mandelbaum}, {Wang}, {Zu}, {White},
  {Henriques}, \& {More}}]{Mandelbaum16}
{Mandelbaum}, R., {Wang}, W., {Zu}, Y., {et~al.} 2016, \mnras, 457, 3200

\bibitem[{{Marulli} {et~al.}(2013)}]{Marulli13}
{Marulli}, F., {et~al.} 2013, \aap, 557, A17

\bibitem[{{Matthews} {et~al.}(2013){Matthews}, {Newman}, {Coil}, {Cooper}, \&
  {Gwyn}}]{Matthews13}
{Matthews}, D.~J., {Newman}, J.~A., {Coil}, A.~L., {Cooper}, M.~C., \& {Gwyn},
  S.~D.~J. 2013, \apjs, 204, 21

\bibitem[{{Meneux} {et~al.}(2006)}]{Meneux06}
{Meneux}, B., {et~al.} 2006, \aap, 452, 387

\bibitem[{{Meneux} {et~al.}(2008)}]{Meneux08}
---. 2008, \aap, 478, 299

\bibitem[{{Meneux} {et~al.}(2009)}]{Meneux09}
---. 2009, \aap, 505, 463

\bibitem[{{More} {et~al.}(2011){More}, {van den Bosch}, {Cacciato}, {Skibba},
  {Mo}, \& {Yang}}]{More11}
{More}, S., {van den Bosch}, F.~C., {Cacciato}, M., {et~al.} 2011, \mnras, 410,
  210

\bibitem[{{Mostek} {et~al.}(2013){Mostek}, {Coil}, {Cooper}, {Davis}, {Newman},
  \& {Weiner}}]{Mostek13}
{Mostek}, N., {Coil}, A.~L., {Cooper}, M., {et~al.} 2013, \apj, 767, 89

\bibitem[{{Moster} {et~al.}(2013){Moster}, {Naab}, \& {White}}]{Moster13}
{Moster}, B.~P., {Naab}, T., \& {White}, S.~D.~M. 2013, \mnras, 428, 3121

\bibitem[{{Moster} {et~al.}(2010){Moster}, {Somerville}, {Maulbetsch}, {van den
  Bosch}, {Macci{\`o}}, {Naab}, \& {Oser}}]{Moster10}
{Moster}, B.~P., {Somerville}, R.~S., {Maulbetsch}, C., {et~al.} 2010, \apj,
  710, 903

\bibitem[{{Moustakas} {et~al.}(2013){Moustakas}, {Coil}, {Aird}, {Blanton},
  {Cool}, {Eisenstein}, {Mendez}, {Wong}, {Zhu}, \& {Arnouts}}]{Moustakas13}
{Moustakas}, J., {Coil}, A.~L., {Aird}, J., {et~al.} 2013, \apj, 767, 50

\bibitem[{{Newman} {et~al.}(2013)}]{Newman13}
{Newman}, J.~A., {et~al.} 2013, \apjs, 208, 5

\bibitem[{{Noeske} {et~al.}(2007)}]{Noeske07}
{Noeske}, K.~G., {et~al.} 2007, \apjl, 660, L47

\bibitem[{{Norberg} {et~al.}(2001)}]{Norberg01}
{Norberg}, P., {et~al.} 2001, MNRAS, 328, 64

\bibitem[{{Norberg} {et~al.}(2002)}]{Norberg02}
---. 2002, \mnras, 332, 827

\bibitem[{{Oliver} {et~al.}(2000)}]{Oliver00}
{Oliver}, S., {et~al.} 2000, \mnras, 316, 749

\bibitem[{{Peacock} \& {Smith}(2000)}]{Peacock00}
{Peacock}, J.~A., \& {Smith}, R.~E. 2000, \mnras, 318, 1144

\bibitem[{{Peebles}(1980)}]{Peebles80}
{Peebles}, P.~J.~E. 1980, {The large-scale structure of the universe}
  (Princeton University Press)

\bibitem[{{Pierre} {et~al.}(2004)}]{Pierre04}
{Pierre}, M., {et~al.} 2004, JCAP, 9, 11

\bibitem[{{Pollo} {et~al.}(2006)}]{Pollo06}
{Pollo}, A., {et~al.} 2006, \aap, 451, 409

\bibitem[{{Prescott} {et~al.}(2011)}]{Prescott11}
{Prescott}, M., {et~al.} 2011, \mnras, 417, 1374

\bibitem[{{Rodr{\'{\i}}guez-Puebla} {et~al.}(2015){Rodr{\'{\i}}guez-Puebla},
  {Avila-Reese}, {Yang}, {Foucaud}, {Drory}, \& {Jing}}]{Rodriguez-Puebla15}
{Rodr{\'{\i}}guez-Puebla}, A., {Avila-Reese}, V., {Yang}, X., {et~al.} 2015,
  \apj, 799, 130

\bibitem[{{Schiminovich} {et~al.}(2007)}]{Schiminovich07}
{Schiminovich}, D., {et~al.} 2007, \apjs, 173, 315

\bibitem[{{Scoville} {et~al.}(2007)}]{Scoville07}
{Scoville}, N., {et~al.} 2007, \apjs, 172, 1

\bibitem[{{Seljak}(2000)}]{Seljak00}
{Seljak}, U. 2000, \mnras, 318, 203

\bibitem[{{Skibba} {et~al.}(2014)}]{Skibba14}
{Skibba}, R.~A., {et~al.} 2014, \apj, 784, 128

\bibitem[{{Skibba} {et~al.}(2015)}]{Skibba15}
---. 2015, \apj, 807, 152

\bibitem[{{Smith} {et~al.}(2003)}]{Smith03}
{Smith}, R.~E., {et~al.} 2003, \mnras, 341, 1311

\bibitem[{{Sobral} {et~al.}(2010){Sobral}, {Best}, {Geach}, {Smail},
  {Cirasuolo}, {Garn}, {Dalton}, \& {Kurk}}]{Sobral10}
{Sobral}, D., {Best}, P.~N., {Geach}, J.~E., {et~al.} 2010, \mnras, 404, 1551

\bibitem[{{Speagle} {et~al.}(2014){Speagle}, {Steinhardt}, {Capak}, \&
  {Silverman}}]{Speagle14}
{Speagle}, J.~S., {Steinhardt}, C.~L., {Capak}, P.~L., \& {Silverman}, J.~D.
  2014, \apjs, 214, 15

\bibitem[{{Strateva} {et~al.}(2001)}]{Strateva01}
{Strateva}, I., {et~al.} 2001, \aj, 122, 1861

\bibitem[{{Velander} \& others.(2014)}]{Velander14}
{Velander}, M., \& others. 2014, \mnras, 437, 2111

\bibitem[{{Wake} {et~al.}(2011)}]{Wake11}
{Wake}, D.~A., {et~al.} 2011, \apj, 728, 46

\bibitem[{{Watson} \& {Conroy}(2013)}]{Watson13}
{Watson}, D.~F., \& {Conroy}, C. 2013, \apj, 772, 139

\bibitem[{{Watson} {et~al.}(2015){Watson}, {Hearin}, {Berlind}, {Becker},
  {Behroozi}, {Skibba}, {Reyes}, {Zentner}, \& {van den Bosch}}]{Watson15}
{Watson}, D.~F., {Hearin}, A.~P., {Berlind}, A.~A., {et~al.} 2015, \mnras, 446,
  651

\bibitem[{{Wetzel} {et~al.}(2012){Wetzel}, {Tinker}, \& {Conroy}}]{Wetzel12}
{Wetzel}, A.~R., {Tinker}, J.~L., \& {Conroy}, C. 2012, \mnras, 424, 232

\bibitem[{{Whitaker} {et~al.}(2014)}]{Whitaker14}
{Whitaker}, K.~E., {et~al.} 2014, \apj, 795, 104

\bibitem[{{Wuyts} {et~al.}(2011)}]{Wuyts11}
{Wuyts}, S., {et~al.} 2011, \apj, 742, 96

\bibitem[{{Zehavi} {et~al.}(2005)}]{Zehavi05}
{Zehavi}, I., {et~al.} 2005, ApJ, 630, 1

\bibitem[{{Zehavi} {et~al.}(2011)}]{Zehavi11}
---. 2011, \apj, 736, 59

\bibitem[{{Zu} \& {Mandelbaum}(2016)}]{Zu16}
{Zu}, Y., \& {Mandelbaum}, R. 2016, \mnras, 457, 4360

\end{thebibliography}

\end{document}